\documentclass[aps,amsfonts,pra,
showpacs,nobibnotes,nofootinbib,
tightenlines]{revtex4}
\input BoxedEPS
 \SetRokickiEPSFSpecial  %% for dvips by Tom Rokicki, for VMS
 \HideDisplacementBoxes
\begin{document}
\newcommand{\beq}{\begin{equation}}
\def\m#1{\mathbf{#1}}
\newcommand{\eeq}{\end{equation}}
\newcommand{\barr}{\begin{eqnarray}}
\newcommand{\earr}{\end{eqnarray}}
\def\figwidth{7.5cm}
\def\qqq{{qqqq\bar q}}
\def\cH{{\cal H}}
\def\cP{{\cal P}}
\def\cD{{\cal D}}
\def\cG{{\cal G}}
\def\cV{{\cal V}}
\def\cF{{\cal F}}
\def\cU{{\cal U}}
\def\cS{{\cal S}}
\def\cO{{\cal O}}
\def\cE{{\cal E}}
\def\bfA{{\bf A}}
\def\bfG{{\bf G}}
\def\bfn{{\bf n}}
\def\bfr{{\bf r}}
\def\bfV{{\bf V}}
\def\bft{{\bf t}}
\def\bfM{{\bf M}}
\def\bfP{{\bf P}}
\def\bra#1{\langle #1 |}
\def\ket#1{| #1 \rangle}
\newcommand{\p}{\partial}
%%%%%%%%%%%%%%%%%%%%%%%%new def%%%%%%%%%%%%%%%%%%%%%%%%%%%
%\newcommand{\ket}[1]{| #1 \rangle}
%\newcommand{\bra}[1]{\langle #1 |}
\def\coltwovector#1#2{\left({#1\atop#2}\right)}
\def\upp{\coltwovector10}
\def\downn{\coltwovector01}
\def\Ord#1{{\cal O}\left( #1\right)}
\def\bmp{\mbox{\boldmath $p$}}
\def\rhobar{\bar{\rho}}
\def\qq{Q\!\!\!\! Q}
\def\rjadd#1{{\bf #1}}
\renewcommand{\Re}{{\rm Re}}
\renewcommand{\Im}{{\rm Im}}
\renewcommand{\theequation}{\thesection.\arabic{equation}}
%%%%%%%%%%%%%%%%%%%%%%%%%%%%%%%%%%%%%%%%%%%%%%%%%%%%%%%%%%
\def\ask{\marginpar{?? ask:  \hfill}}
\def\fin{\marginpar{fill in ... \hfill}}
\def\note{\marginpar{note \hfill}}
\def\check{\marginpar{check \hfill}}
\def\discuss{\marginpar{discuss \hfill}}
%%%%%%%%%%%%%%%%%%%%%%%%%%%%%%%%%
%Title of paper
\title{{\Large Exotica}\footnote{These notes
are a distillation of talks presented at QM2004, Berkeley, CA, January
2004; Multiquark Hadrons: Four, Five and More?, Kyoto, Japan, February
2004; Continuous Advances in QCD 2004, Minneapolis, MN, May 2004; the
JLab Annual Users Meeting, June 2004; BEACH'04, Chicago, IL, June
2004; and The Workshop on Strings \& QCD, Trento, Italy, July 2004} }
\author{R.~L.~Jaffe}\email{jaffe@mit.edu} \affiliation{Center for
Theoretical Physics \\
Laboratory for
   Nuclear Science
   and Department of Physics \\
Massachusetts Institute
  of Technology \\ Cambridge, MA 02139, USA}
 
%\date{\today}
\begin{abstract}
	\noindent The first evidence for 
	Quantum Chromodynamics (QCD), the theory of the strong 
	interactions, came from the systematics of baryon and meson 
	spectroscopy.  An important early observation was the apparent absence 
	of exotics, baryons requiring more than three quarks or mesons 
	requiring more than $q\bar q$.  Years later, QCD is well 
	established, hadron spectroscopy has been relatively inactive, but 
	the absence of exotics remains poorly understood.  
	The recent observation of narrow, prominent exotic
	baryons has stirred up new interest in hadron spectroscopy.  At present
	the experimental situation is confused; so is theory.  The recent
	discoveries are striking.  So too is the complete \emph{absence}
	of exotic mesons, and, except for the recent discoveries, of
	exotic baryons as well.  Whether or not the new states are
	confirmed, the way we look at complicated states of confined
	quarks and gluons has changed.  Perhaps the most lasting result,
	and the one emphasized in these notes, is a new appreciation
	for the role of diquark correlations in QCD.  
\end{abstract}
\vspace*{-\bigskipamount}  \preprint{MIT-CTP-3538} 
\maketitle
\section{Introduction}
\setcounter{equation}{0} 

There is no doubt that Quantum Chromodynamics is the correct theory of
the strong interactions.  It has been tested quantitatively in
hundreds of experiments at high momentum transfer, where asymptotic
freedom justifies the use of perturbation theory\cite{af}.  Hadrons
are clearly bound states of quarks held together by gluon mediated,
non-Abelian gauge interactions.  After many years, however, a
quantitative and predictive theory of confined states of quarks and
gluons still eludes us.  This is particularly true for the light, $u$,
$d$, and $s$, quarks, where non-relativistic approximations fail. 
Hadron spectroscopy is interesting in its own right, but also because
it is a laboratory in which to explore the dynamics of an unbroken
gauge interaction with a non-trivial ground state, a model for other
unsolved problems in high energy physics.  Also, the spectrum of
hadrons shows many qualitative regularities that do not follow simply
from the symmetries of QCD, and invite a deeper understanding.  The
one which figures in the present discussion is that all known hadrons
can be described as bound states of $qqq$ or $q\bar q$.  Because QCD
conserves the number of quarks of each flavor ($u$, $d$, $s$,\ldots),
hadrons can be labeled by their minimum, or \textit{valence}, quark
content.  Thus, for example, the conserved quantum numbers of the
$\Lambda$ hyperon require that include $uds$.  QCD can augment this
with flavor neutral pairs ($u\bar u$, $d\bar d$, \textit{etc.}) or
gluons, but only three quarks are required to account for the
conserved quantum numbers of the $\Lambda$.  Likewise, the $K^{+}$
meson must include at least $u\bar s$.  Hadrons whose quantum numbers
require a valence quark content beyond $qqq$ or $q\bar q$ are termed
``exotics''.  The classic example is a baryon with positive
strangenss, a $Z^{*}$ as it is known, with valence quark content
$uudd\bar s$.\footnote{Accidentally, strangeness was assigned to
hadrons in the 1950's in a way such that the $s$-quark ended up with
negative strangeness.  A similar choice by B.~Franklin assigned the
electron negative electric charge.}

The absence of exotics is one of the most obvious features of QCD. In
the early years experimenters searched hard for baryons that cannot be
made of three quarks or mesons that cannot be made of $q\bar
q$.\footnote{A small number of mesons whose spin, parity and charge
conjugation are forbidden in the \emph{non-relativistic} quark model
are also often termed ``exotics''.  These can also be $\bar q q
g$(gluon) states or even relativistic $\bar q q$ bound states.  They
will not figure in these notes.  Also, I will not discuss heavy quark
mesons with unexpected masses.} Exotic mesons seemed entirely absent. 
Controversial signals for exotic baryons known as $Z^{*}$s came, and
usually went, never rising to a level of certainty sufficient for the
Particle Data Group's tables\cite{pdg1982,z*disc}.  In its 1988 review
the Particle Data Group officially put the subject to
sleep\cite{pdg1988}:
\begin{quote}
	``The general prejudice against baryons not made of three quarks
	and the lack of any experimental activity in this area make it
	likely that it will be another 15 years before the issue is
	decided.''
\end{quote}
After that, the subject of exotic baryons did not receive much
attention except from a small band of theorists motivated by the
predictions of chiral soliton models\cite{man,chem,
bied,pras,wall,dpp,weigel}.  Then, in January of 2003 evidence was
reported of a very narrow baryon with strangeness one and charge one,
of mass $\approx 1540$ MeV, now dubbed the $\Theta^{+}$, with minimum
quark content $uudd\bar s$\cite{spring8}.  The first experiment was
followed by many other sightings\cite{thetasightings} and by evidence
for other exotics: a strangeness minus two, charge minus two particle
now officially named the $\Phi^{--}$ by the PDG, with minimum quark
content $ddss\bar u$\cite{na49} at 1860 MeV , and an as-yet nameless
charm exotic $(uudd\bar c)$\cite{h1} at 3099 MeV. Theorists, myself
included, descended upon these reports and tried to extract dynamical
insight into QCD\cite{theory}.  Other experimental groups began
searches for the $\Theta^{+}$ and its friends.  As time has passed the
situation has become more, rather than less, confusing\cite{trilling}:
several experiments have now reported negative results in searches for
the $\Theta^{+}$\cite{dzierba}; no one has confirmed either the
$\Phi^{--}(1860)$ or the $uudd\bar c$(3099); and theorists have yet to
find a compelling (to me at least) explanation for the low mass or
narrow width of the $\Theta^{+}$.

The existence of the $\Theta^{+}$ is a question for experimenters. 
Theorists simply do not know enough about QCD to predict without doubt
whether a light, narrow exotic baryon exists.  Whether or not the
$\Theta^{+}$ survives, it is clear that exotics are very rare in QCD.
Perhaps they are entirely absent.  This remarkable feature of QCD is
often forgotten when exotic candidates are discussed.  The existence
of a handful of exotics has to be understood in a framework that also
explains their overall rarity.  Along the same line, the \emph{aufbau}
principle of QCD differs dramatically from that of atoms and nuclei:
to make more atoms add electrons, to make more nuclei, add neutrons
and protons.  However in QCD the spectrum --- with the possible
exception of a few states like the $\Theta^{+}$ --- seems to stop at
$qqq$ and $q\bar q$.

Thinking about this problem in light of early work on multiquark
correlations in QCD \cite{rjmulti}, Frank Wilczek and
I\cite{jw}\footnote{Closely related ideas have been explored by
Nussinov\cite{sn} and by Karliner and Lipkin\cite{kl}.} began to
re-examine the role of diquark correlations in QCD. Diquarks are not
new; they have been championed by a small group of QCD theorists for
several decades\cite{firstdiquarks,diquarks}.  We already
knew\cite{rjmulti,erice} that diquark correlations can naturally
explain the general absence of exotics and predict a supernumerary
nonet of scalar mesons which seems to exist.  We quickly learned that
they can rather naturally accommodate exotics like the $\Theta^{+}$. 
They also seem to be important in dense quark matter\cite{colorsuper},
to influence quark distribution\cite{closethomas} and fragmentation
functions, and to explain the systematics of non-leptonic weak decays
of light quark baryons and mesons\cite{neubert}.  Whether or not the
$\Theta^{+}$ survives, diquarks are here to stay.

In the first part of this paper, after looking briefly at the history
of exotics, I assume that the $\Theta^{+}$ exists, and see how well it
fits with other features of light quark spectroscopy.  I will take a
look at the $\Theta^{+}$ from several perspectives: general scattering
theory, large $N_{\rm c}$, chiral soliton models, quark models, and
lattice calculations.  In general these exercises raise more questions
than they answer.  In brief: A light, narrow exotic is inconvenient
but not impossible for QCD spectroscopy.

Thinking about the $\Theta^{+}$ in terms of quarks leads one naturally
to study quark correlations, and especially diquarks.  So the later
sections of the paper are devoted to diquarks.  I define them
carefully and review some of the evidence that they are important in
QCD. Next I describe how diquark correlations in hadrons can explain
qualitatively most of the puzzles of exotic hadron spectroscopy:
first, why exotics are so rare in QCD; next, why there is an extra
nonet of scalar mesons; third, why an exotic baryon antidecuplet
containing the $\Theta^{+}$ would be the only prominent baryon exotic;
fourth, why non-strange systems of 6, 9, 12, \ldots quarks form nuclei
not single hadrons; and finally why the $H$ dibaryon ($uuddss$) might
not be as bound as simple estimates suggest\cite{rljh}. 
``Qualitatively'' is an important modifier, however: like all quark
model ideas, this one lacks a quantitative foundation.  Perhaps
lattice QCD studies in the not-too-distant future can confirm some
aspects of the analysis, but the need for a systematic and predictive
phenomenological framework for QCD spectroscopy has never been
greater.
 
This paper is not a review, but instead an ideosyncratic overview and
introduction aimed at readers who may not already be familiar with the
subject.\footnote{They has been tested on string theorists, for
example.} I focus principally on quark-based dynamics and on diquark
correlations which, I believe, are strongly motivated by other
features of hadron phenomenology.  Hundreds of theorists have written
on the subject of the $\Theta$ and other exotic baryons from different
perspectives.  Presentations of other points of view can be found in
Refs.~\cite{diak, others}.  The paper is not very technical.  A few
subsections sections are more detailed, and some contain previously
unpublished material (see especially, \S IV.B and D).  Since there is
no ``Theory of the Spectrum'' in QCD, detailed calculations do not
seem warranted.  I will concentrate on the qualitative features of
models that can provide a guide to the study of exotics.

Here is an outline: 
\begin{enumerate}
	\item[II.] History
	\begin{enumerate}
		\item[A.]	The  absence of exotics
		\item[B.] Exotic sightings since January 2003
	\end{enumerate}
	\item[III.] Theoretical perspectives
	\begin{enumerate}
		\item[A.] Insights from scattering theory.
		\item[B.] Large $N_{\rm c}$ and chiral soliton models
		\begin{itemize}
			\item[1.] Large $N_{\rm c}$
			\item[2.] Chiral Soliton Models
		\end{itemize}
		\item[C.] Quark models.
		\begin{itemize}
			\item[1.] General features of an uncorrelated quark model
			\item[2.] Quark model ``states'' and scattering
			\item[3.] Pentaquarks in the uncorrelated quark model
		\end{itemize}	
		\item[D.] Early lattice results
	\end{enumerate}
	\item[IV.] Diquarks
	\begin{enumerate}
		\item[A.] Introducing diquarks
		\item[B.] Characterizing diquarks
		\item[C.] Phenomenological evidence for diquarks
		\item[D.] Diquarks and higher twist
	\end{enumerate}
	\item[V.] Diquarks and Exotics
	\begin{enumerate}
		\item[A.] An overview
		\item[B.] The scalar mesons
		\item[C.] Pentaquarks from diquarks I:   The general idea
		\item[D.] Pentaquarks from diquarks II:  A more detailed look at 
the positive 
		parity octet and antidecuplet
		\begin{itemize}
			\item[1.] Flavor $SU(3)$ violation and mass relations
			\item[2.] Isospin and SU(3) selection rules
		\end{itemize}
		\item[E.] Pentaquarks from diquarks III: Charm and bottom
		analogues \item[F.] A paradigm for spectroscopy
	\end{enumerate}
	\item[VI.] Conclusions
\end{enumerate}

\section{History}
\setcounter{equation}{0} 
\subsection{The   absence of exotics}

Most talks on the $\Theta^+$ begin by showing the experimental 
evidence
reported in the past two years.  I would like to strike a different
note by beginning with a brief look at evidence of the absence of
exotics.  Spectroscopy was at the cutting edge of high energy physics
in the `60's and `70's.  A great deal of effort and sophisticated
analysis was brought to bear on the study of the hadron spectrum, and
the conclusions remain important.

The fact that all known hadrons made of light, $u$, $d$, and $s$
quarks can be classified in $SU(3)_{\rm f}$ representations that can 
be
formed from $\bar q q$ or $qqq$ played an essential role in the
discovery of quarks.  Gell-Mann mentions it prominently in his first
paper on quarks in 1964\cite{mgm}:
\begin{quote}
	``Baryons can be constructed from quarks by using the combinations
	$(qqq)$, $(\qqq)$, {\it etc\/}, while mesons are made out
	of $(q\bar q)$, $(qq\bar q\bar q)$, {\it etc\/}.  It is assuming
	[{\it sic}] that the lowest baryon configuration $(qqq)$ gives
	just the representations $\mathbf {1}$, $\mathbf{ 8}$, and
	$\mathbf{ 10}$ that have been observed, while the lowest meson
	configuration $(q\bar q)$ similarly gives just $\mathbf{1}$ and
	$\mathbf{ 8}$.''
\end{quote}
In the decades that followed, many excellent experimental groups
studied meson-baryon and meson-meson scattering, and extracted the
masses and widths of meson and baryon resonances.  Resonances were
discovered in nearly all non-exotic meson and baryon channels, but no
prominent exotics were found.

\begin{figure}
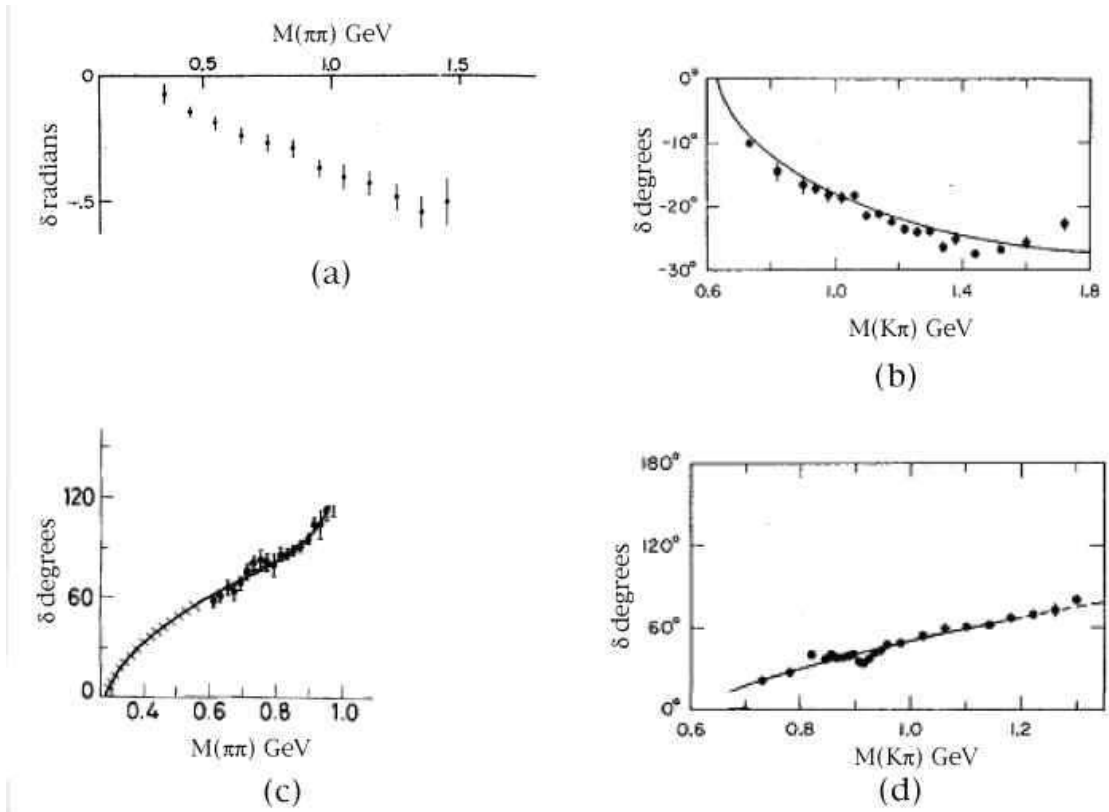

\begin{center}
\BoxedEPSF{phases-.epsf scaled 600} \caption{Meson-meson $s$-wave
scattering phase shifts as they were known in the 1970's.  (a)
$\pi\pi$ in the exotic $I=2\ (\pi^{+}\pi^{+})$ channel\cite{meson1};
(b) $\pi K$ in the exotic $I=3/2\ (\pi^{+}K^{+})$
channel\cite{meson2}; (c) $\pi\pi$ in the non-exotic $I=0$
channel\cite{meson1}; (d) $\pi K$ in the non-exotic $I=1/2$
channel\cite{meson2}.  The exotic channels show weak repulsion (slowly
falling phases).  The non-exotic channels show strong attraction
(steadily rising phases).}
\label{mesonphases}
\end{center}
\end{figure}

Figure \ref{mesonphases} shows the $\pi\pi$ and $K\pi$ phase shifts in
the $s$-wave for the exotic ($\pi^{+}\pi^{+}$ and $\pi^{+}K^{+}$) and
non-exotic channels as they were known in the late
1970's\cite{meson1,meson2}.  The non-exotic channels show positive,
slowly increasing phases which we now associate with the scalar
mesons, $f_{0}(600)$ and $\kappa(800)$.  The exotic channels show
small, negative, slowly falling phases characteristic of a weak
repulsive interaction.  Similar behavior was observed in all exotic
channels that were studied.

\begin{figure}
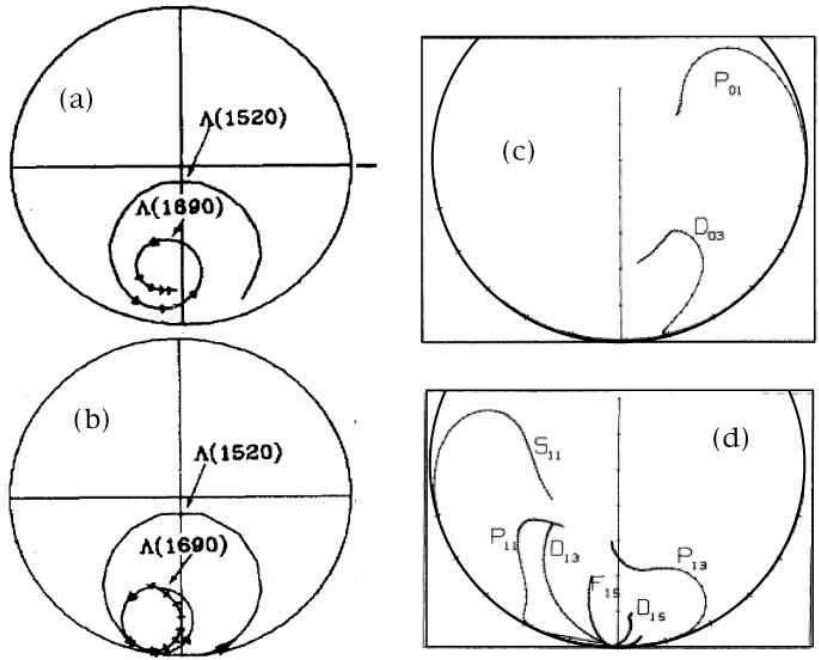
 
	\begin{center}
	\BoxedEPSF{argands-.epsf scaled 450} \caption{Argand diagrams for
	(a), (b) $I=0$ $\overline KN$ scattering from two different
	experiments in the non-exotic $D_{03}$ wave, and $KN$ scattering
	for (c) $I=0$, and (d) $I=1$ in various partial waves.  The
	partial wave notation is $L_{I,2J}$.  (a) and (b) show evidence
	for prominent resonances, and the data from two different
	experiments\cite{pdg84,expt1,expt2} agree with one another.  In
	the exotic channels ((c) and (d)) there is more and better data on
	$I=1$ because the $K^{+}p$ channel is more easily studied than
	$K^{+}n$.  The quality of the figures\cite{arndtetal} is not
	sufficient to see clearly the tick marks on the curves at
	intervals of 50 MeV in beam energy.  The non-exotic resonances are
	relatively narrow.  All the exotic effects change slowly with
	energy.  The $[KN]^{I=0}$ $s$-wave ($S_{01}$) is not shown in (c)
	but is weakly repulsive.  The broad counterclockwise motions in
	the $P_{01}$, $D_{03}$, $P_{13}$, and $D_{15}$ partial waves were
	interpreted as evidence for broad, inelastic resonances near 1830,
	1790, 1810, and 2070 MeV respectively.  See Ref.~\cite{MJ2003} for
	further discussion.  At an elastic resonance, like the $\Theta^+$,
	the argand amplitude should execute a complete loop on the outer
	(unitarity) circle.}
	\label{argands} 
	\end{center}
\end{figure}

The situation among the baryons has always been more complicated. 
More is known because meson-baryon scattering is easier to study than
meson-meson.  During the 1970's and `80's there were candidates for
broad and/or inelastic exotic $KN$ resonances --- they were the
subjects of the 1982 and 1988 PDG reviews quoted
above\cite{pdg1982,pdg1988}.  The history and status of these states
has recently been reviewed by Jennings and Maltman\cite{MJ2003}. 
Figure \ref{argands} shows the Argand diagrams for elastic scattering
in one non-exotic $\overline KN$ channel, and in the exotic $KN$
channels ($K^{+}p$ with isospin one, and $K^{0}p/K^{+}n$ with isospin
zero), corresponding to quark content $qqqq\bar s$, where $q$ denotes
a light, $u$ or $d$, quark \cite{arndtetal}.  The non-exotic channel
shown for comparison is the $\overline KN$ $d$-wave with
$J^{\Pi}=3/2^{-}$, a channel with two well established resonances.  In
Fig.~\ref{argands} the Argand amplitude,
\begin{equation}	
f_{\ell}(k)=\frac{i}{2}-\frac{i}{2}\eta_{\ell}(k)e^{2i\delta_{\ell}(k)}
\label{argand}
\end{equation}
is plotted parametrically as a function of kaon energy in the nucleon
rest frame. 
If the scattering is elastic, then $\eta_{\ell}=1$ and
$f_{\ell}(k)$ must lie on the unitarity circle,
$|f_{\ell}(k)-\frac{i}{2}|=\frac{1}{2}$.  When $\eta_{\ell}=1$ the
elastic cross section in each partial wave is given simply by
\begin{equation}
\sigma_{\ell}(k) = \frac{4\pi(2\ell+1)}{k^{2}}\sin^{2}\delta_{\ell}(k)
\label{sigma}
\end{equation}
An elastic resonance gives rise to a rapid increase of
$\delta_{\ell}(k)$ by $\sim 2\pi$ and appears as a counterclockwise
circle of radius 1/2 in $f_{\ell}(k)$.  If there is no background
phase, {\it ie\/} if $\delta_{\ell}\approx 0$ just before the onset
of the resonance, then at the peak of the resonance
$\delta_{\ell}=\pi/2$ and $\sigma_{\ell}=4\pi(2\ell+1)/k^{2}$.  A
significant background phase can alter the shape of a narrow
resonance\cite{shape}, but because of unitarity it cannot reduce the
magnitude of the effect on the cross section.  The only way to miss an
elastic resonance is if its width is significantly smaller than
experimental resolution.  At higher energies more channels open,
scattering becomes inelastic, and resonances are associated with less
pronounced counterclockwise arcs.  The non-exotic $\overline KN$
channels show many clear resonances at low energy.  The exotic
channels show none.  When the PDG wrote its 1988 review, the closest
thing to an exotic was the broad, inelastic counterclockwise motion in
the $P_{01}$ partial wave shown in Figure~\ref{argands}(c).

The zeroth order summary prior to January 2003 was simple: no exotic
mesons or baryons.  In fact the only striking anomaly in low energy
scattering was the existence of a supernumerary ({\it ie\/} not
expected in the quark model) nonet of scalar, ($J^{\Pi}=0^{+}$) mesons
with masses below 1 GeV: the $f_{0}(600)$, $\kappa(800)$,
$f_{0}(980)$, and $a_{0}(980)$, about which more later.

When the $\Theta^+$ was first reported, several groups re-examined the
old $KN$ scattering data and interpreted the absence of any structure
near 1540 MeV as an upper limit on the width of the
$\Theta^+$\cite{sn,ct,aws,meiss}.  The limits range from 0.8 MeV 
through
``a few'' MeV. It is important to remember that these are not
sightings of a narrow $\Theta^+$, rather they are reports of negative
results expressed as an upper limit on the width of the $\Theta^+$.

\subsection{Exotic sightings since January 2003}

\begin{figure}
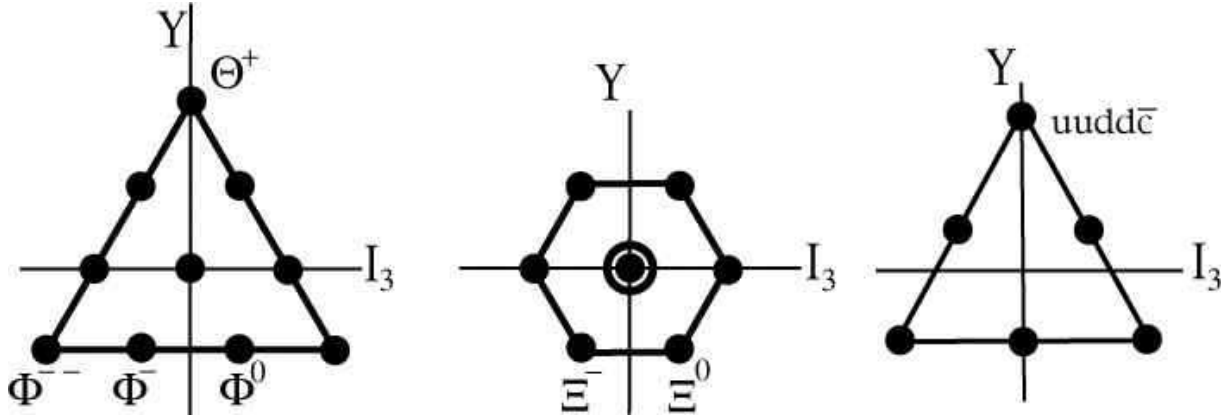
 
	\begin{center}
	\BoxedEPSF{10barand8-.epsf scaled 650} \caption {Simplest possible
	$SU(3)_{\rm f}$ representations for the exotic candidate exotic
	baryons: (a) the $\m{\overline{10}}$; (b) the $\m{8}$; and (c) the
	$\m{\overline{6}}$.}
	\label{10barand8} 
	\end{center}
\end{figure}

Space and time do not permit me to present and review all the reports
of exotics since January of 2003.  Instead I have tried to summarize
the situation in two tables.  The first, Table \ref{theta}, is derived
from one presented by T.~Nakano\cite{nakano}, reporting sightings of
the $\Theta^+$.  The second, Table \ref{sightings}, is a summary of 
the
properties of the reported states.  The baryons in
Table~\ref{sightings} can be classified in the $\m{\overline{10}}$ or
$\m{8}$ and $\m{\overline{6}}$ representations of $SU(3)_{\rm f}$ as 
shown
in Fig.~\ref{10barand8}.\footnote{$SU(3)$-flavor representations are
denoted by their dimension in boldface.  Irreps of other symmetry
groups like $SU(3)$-color, or the $SU(6)$ symmetries built from
flavor$\times$spin or color$\times$spin are distinguished by
appropriate subscripts.  Occasionally a subscript ``$f$'' is added to
an $SU(3)$-flavor representation for clarity.}

Although they could be in higher representations, the
$\m{\overline{10}}$ is the simplest that can accomodate both the
$\Theta^{+}$ and the $\Phi^{--}$.  I have not attempted to summarize
the searches that fail to see the $\Theta^+$ or the other newÁ
exotics.  These require a careful discussion, which can be found, for
example in Ref.~\cite{dzierba}.

\begin{table}[htdp]
\caption{Summary of sightings of the $\Theta^{+}$\cite{nakano}}
\begin{center}
\begin{tabular}{|c|c|c|c|c|}
\hline
Experiment & Reaction&Mass&Width&$\sigma_{\rm std}$\\
\hline
LEPS &\quad $\gamma C\to K^+K^{-}\ X$&$1540~\pm~10$&$<25$&4.6\\
\hline
Diana &\quad $K^+Xe \to K^0p X$&$1539~\pm~2$&$<9$&4.4\\
\hline
CLAS &\quad $\gamma d \to K^+ K^- p(n)$&$1542~\pm~5$&$<21$&5.2\\
\hline
SAPHIR& \quad $\gamma p \to K^+ K^0 (n)$&$1540~\pm~6$&$<25$&4.8\\
\hline
ITEP &\quad $vA \to K^0p X $&$ 1533~\pm 5$&$<20$&6.7\\ 
\hline
CLAS &\quad $\gamma p \to \pi^+ K^- K^+ (n)$&$1555~\pm~10$&$<26$&7.8\\
\hline
HERMES& \quad $e^+ d \to K^0p X$&$1528~\pm~3$&$13~\pm~9$& $\sim$5\\
\hline
ZEUS &\quad $e^+p \to e^\prime K^0 p 
X$&$1522~\pm~3$&$8~\pm~4$&$\sim$5\\ 
\hline
COSY &\quad $pp \to K^0 p \sum^+$&$1530~\pm~5$&$<18$&$4-6$\\
\hline
\end{tabular}
\end{center}
\label{theta}
\end{table}%

\begin{table}[htdp]
	\caption{Properties of reported exotic baryons and related states.}
\begin{center}
\begin{tabular}{|c|c|c|c|c|c|c|}
\hline
Name & Mass (MeV) &Width (MeV)&Spin$^{\rm parity}$&Isospin& Decays & 
Minimal $SU(3)_{f}$ Irrep\\
\hline
$\Theta^{+}$ &1520---1540&\ $<1^{[1]},\  <6-10^{[2]}$\ & $1/2^{? 
[3]}$& 0 &
	$K^{+}n,\ K_{S}p$& $\m{\overline{10}}$\\
\hline
$\Phi^{--}$ &1860&$< 18$&$?$& $\ge 3/2$ &$\Xi^{-}(1320)\pi^{-}$&$\m{\overline{10}}$\\
\hline
$\Phi^{0}/\Xi^{0 [4]}$  &1860&$< 18$&?&$\ge 
1/2$&$\Xi^{-}(1320)\pi^{+}$&
$ \m{\overline{10}}$ if $\Phi$,\ $\m{8}$ if $\Xi$ \\
\hline
$\Phi^{-}/\Xi^{- [4]}$ &1855&$<18$&?&$\ge
1/2^{[5]}$&$\Xi^{*0}(1530)\pi^{-}$& 
$ \m{\overline{10}}$ if $\Phi$,\ $\m{8}$ if $\Xi$\\
\hline
$\{uudd\bar c\}$&3099&$ < 12 $&$?$&$\ge 0$&$pD^{*-}\ \&\ \bar p 
D^{*+}$&$\m{\overline{6}}$\\
\hline
\end{tabular}
\end{center}  
\label{sightings}
\end{table}%
\vspace*{-.2in}
\qquad\quad\quad
\parbox[t]{.7\textwidth}{\raggedright \noindent{\scriptsize $^{[1]}$ From analysis of $KN$ scattering; \\  
$^{[2]}$ from direct detection of $\Theta$; \\ 
$^{[3]}$ Weak evidence for $J=1/2$, no information on parity; \\  
$^{[4]}$ $\Phi\,\  \mbox{if} I=3/2,  \Xi\,\  \mbox{if} I=1/2$; \\ 
$^{[5]}$ $I=1/2\, \mbox{favored}\, \mbox{if decay to}   \Xi^{*0}(1530)\ 
 \mbox{is correct.}$}}

\vspace*{.2in}

There are several puzzling aspects of the data: the variation of the
$\Theta^+$ mass and the claims of HERMES and ZEUS to have measured a
non-vanishing width, for example.  The interested reader should
consult the talks by Nakano\cite{nakano} and Dzierba\cite{dzierba} and
other presentations at QNP2004.  Here are some questions and
observations about the data\ldots
\begin{itemize}
	\item Zeus sees the $\Theta^+$ in the current fragmentation region
	in deep inelastic positron scattering.  Particles produced in this
	region are fragments of the struck quark.  Standard factorization
	arguments imply that the fragmentation function $D_{q/\Theta}(z)$
	cannot be zero if the $\Theta^+$ is seen in this experiment. 
	Among those that \emph{have not} seen the $\Theta^+$ are several
	collider experiments in which particle production also occurs
	through quark fragmentation.  Especially relevant are the
	$e^{+}e^{-}$ colliders ({\it eg.\/} Aleph, BABAR) where all
	hadrons are fragments of high momentum quarks \cite{ichep}.  It
	would be interesting to know if these negative results can be
	reconciled with the sighting at Zeus.  Initial reports suggest
	that they are not compatible\cite{babarno}.
	
	\item Many unusual hadrons are formed rather abundantly by quark
	fragmentation in $e^{+}e^{-}$ annihilation.  Examples taken from
	the Durham data base\cite{ddb} include the $\Omega^{-}$ and
	$f_{0}(980)$, which is probably dominantly a $\bar q\bar qqq$
	state.  For example $\sigma(f_{0}(980))/\sigma(\rho)\approx 0.1 -
	0.2$ over a range of $z$.  It would be very helpful if the
	collider experiments would phrase their failure to see the
	$\Theta^+$ as a limit on the $q\to\Theta$ fragmentation function so
	we could compare it with the fragmentation functions of other
	hadrons\cite{babarreport}.
	
	\item The report from COSY/TOF of a sighting of the $\Theta^+$ in
	the reaction $pp\to\Sigma^{+}\Theta^{+}$ is particularly
	interesting.  The $\Sigma^{+}$ is known to couple to $K^{0}p$ and
	the $\Theta^+$ has been reported in $K_{S}p$ invariant mass plots. 
	Therefore it is possible to estimate the cross section for the
	reaction $pp\to\Sigma^{+}\Theta^{+}$ mediated by $K^{0}$ exchange. 
	Theorists have examined the COSY/TOF data and concluded that the
	cross section is roughly consistent with  
	expectations\cite{cosytheory}.
\end{itemize}
	
There are many predominantly experimental issues that I have not
covered here: proposals to measure the parity of the $\Theta^+$, 
limits
on the production of other exotics like a $\Theta^{++}$, and reports
of other bumps with the quantum numbers of the $\Theta^+$ at higher
mass, to name only a few.

\section{Theoretical perspectives}
\setcounter{equation}{0}   

\subsection{Insight from scattering theory}

\begin{figure}
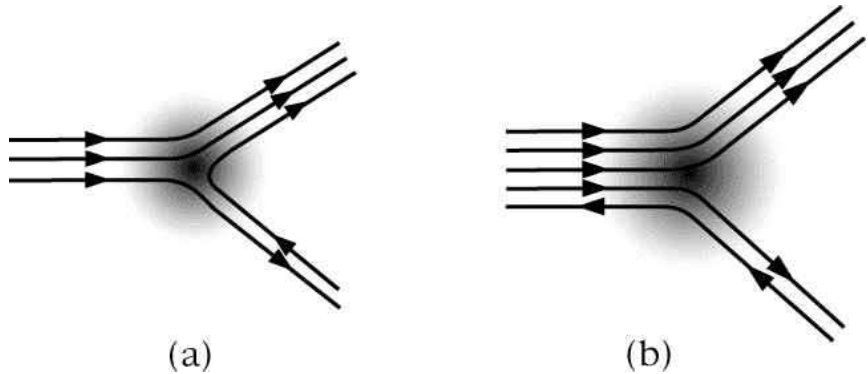

\begin{center}
\BoxedEPSF{quarklines-.epsf scaled 500} \caption{Quark-line diagrams,
(a) for decay of a non-exotic, $qqq$, resonance into a meson and
baryon, requires quark pair creation; (b) for decay of an exotic 
$qqqq\bar q$ resonance is not suppressed by quark pair creation.}
\label{quarklines}
\end{center}
\end{figure}

The small width of the $\Theta^+$ poses a challenge for any
theoretical interpretation.  It is clear that the width is small, but
small compared to what?  The $\Theta^+$ is unique among hadrons in
that its valence quark configuration, $uudd\bar s$, already contains
all the quarks needed for it to decay into $KN$.  Non-exotic hadrons
like the $\rho(770)$ or $\Lambda(1520)$ in their valence quark
configuration can only couple to their decay channels ($\pi\pi$ for
the former, $\overline K N $ for the latter) by creating quark pairs
(see Figure~\ref{quarklines})
The suppression of quark pair creation, known as the Okubo, Zweig,
Iizuki (OZI) Rule\cite{OZI}, is often invoked as an explanation for
the relative narrowness of hadronic resonances.  Some other, as yet
unknown mechanism would have to be responsible for the narrowness of
the $\Theta^+$\cite{speculations}.

General principles of scattering theory allow one to get at least a
qualitative answer to the question: ``How unusual is the width of the
$\Theta^+$?''\cite{jj}.  The $\Theta^+$ appears as an elastic 
resonance in
$KN$ scattering at low energy.  The center of mass momentum is low
enough, $k\approx 270$ MeV, that the motion is arguably
non-relativistic, $\beta_{p}^{2} \approx 0.08$ and
$\beta_{K}^{2}\approx 0.30$ and the Schr\"odinger equation can be used
to examine the scattering.  There is only one open channel (with a
definite isospin), simplifying the problem even further.

There are two ways to make a resonance in low energy scattering,
either (a) the resonance is generated by the forces between the
scattering particles, or (b) it exists in another channel, which is
closed (or confined), and couples to the scattering channel by some
interaction.  The former are the standard resonances of low energy
potential scattering, described in any book on quantum mechanics.  The
latter are ``CDD poles'' that have to be added by hand into the
$S$-matrix\cite{cdd,books}.  Potential scattering resonances (case
(a)) are generated by the interplay between attraction due to
interparticle forces and repulsion, usually due to the angular
momentum barrier.  These resonances subside into the continuum as the
interaction is turned off.  The classic example of a CDD pole (case
(b)) is the $\pi^{-}$-pole in $e^{-}\overline\nu_{e}$ scattering.  It
is not generated by the forces between the electron and antineutrino. 
Rather than disappearing, it decouples, {\it ie\/} its width goes to
zero, as the interaction is turned off.  Another important example of
a CDD pole is a bound state in a closed or confined channel that
couples to a scattering channel by an interaction.  In the case of a
$qqq$ baryon coupling to the meson($\bar qq$)-baryons($qqq$)
continuum, quark pair creation is the interaction.  Thus we should
expect typical baryon resonances to appear as CDD poles in
meson-baryon scattering, not generated by the well known
phenomenological meson-baryon potentials.  The $\Theta^+$ is unusual:
Because its valence quark content, $uudd\bar s$, is the same as the
valence quark content of $KN$, the possibility that it arises from the
$KN$ potential cannot be excluded {\it a priori\/} and has to be
analyzed.

For potential scattering we assume an attractive interaction with
range, $b$, and depth, $V_{0}$.\footnote{Potentials with repulsion at
long distances and attraction at short distances would yield different
results, but seem unnatural.} Keeping the resonance energy fixed at
$M_{\Theta}$=1540 MeV, we obtain a relation between the range and the
width, $\Gamma(\Theta)$, for each value of the orbital angular
momentum $\ell$.  $\ell=0$ can be excluded immediately: there are no
$s$-wave resonances in an attractive potential.  The $p$-wave is
excluded because the range would have to be unnaturally short,
$b\lesssim 0.05$ fm to obtain $\Gamma(\Theta)\lesssim 5$ MeV. Even the
$d$-wave is marginal.  It seems that the $\Theta^+$, narrow as it is,
cannot originate in the $KN$ forces, unless it has angular momentum
much larger than generally supposed or those forces are bizarre.

\begin{figure}
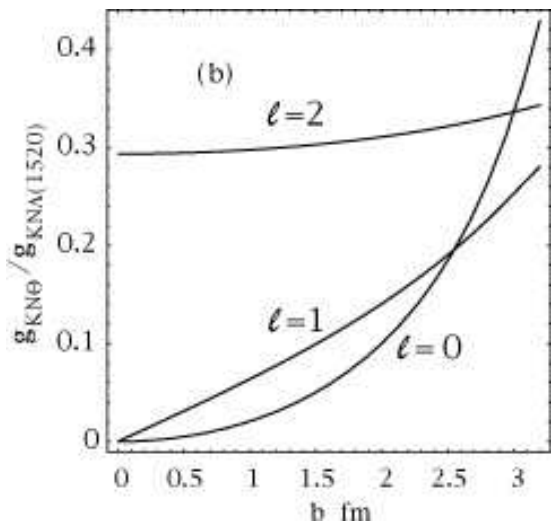

\begin{center}
\BoxedEPSF{cdd-.epsf scaled 850} \caption{Ratio of the channel couplings
for $KN\Theta$ and $\overline KN\Lambda(1520)$, for $\ell=0,1,2$,
assuming that the width of the $\Theta^+$ is 1 MeV.  The ratio scales 
like $\sqrt{\Gamma_{\Theta}}$.}
\label{cdd}
\end{center}
\end{figure}

It is always possible to introduce a CDD pole at 1540 MeV and couple
it weakly enough to give as narrow a width for the $\Theta^+$ as
required.  However, Nature has given us a ``standard'' quark model
resonance, the $\Lambda (1520)$, with valence quark content $uds$ at
nearly the same mass.  This enables us to compare the underlying
$KN\Theta$ coupling, $g_{KN\Theta}$ to the underlying $\overline
KN\Lambda(1520)$ coupling, $g_{\overline KN\Lambda(1520)}$.  The
$\Lambda(1520)$ has $J^{\Pi}=3/2^{-}$ and therefore appears in the
$\overline KN$ $d$-wave.  Its partial width into $\overline KN$ is
$\approx$ 7 MeV. A coupled channel analysis described in
Ref.~\cite{jaffejain} is summarized in Fig.~\ref{cdd},
where the ratio of the $g_{KN\Theta}/g_{\overline KN\Lambda(1520)}$ is
plotted as a function of the range of the interaction that couples the
state to the $KN/\overline KN$ channel for $\ell=0$, 1, and 2.  Values
of $g_{KN\Theta}/g_{\overline KN\Lambda(1520)}$ as small as unity
would already be surprising.  After all, the $\overline KN$ coupling
of the $\Lambda(1520)$ is suppressed by the OZI rule and the
$\Theta^+$ coupling to $KN$ is not.  For $\ell=0$ this ratio would
have to be $\approx 0.02$ to obtain $\Gamma(\Theta)\approx 1$ MeV. For
$\ell=1$ a suppression of $\approx 0.06$ is necessary.  Even for
$\ell\ge 2$ the $KN$ coupling of the $\Theta^+$ would still have to be
less than the the $\overline KN$ coupling of the $\Lambda(1520)$ by a
factor of order the square root of the ratio of their widths to $KN$,
{\it ie\/} by $\approx 0.3$.

The lesson of this exercise is qualitative: If the $\Theta^+$ appears
in a low partial wave ($\ell=0$ or 1), schemes which hope to produce
it from $KN$ forces seem doomed from the start; schemes which
introduce confined channels (quark models are an example, where
reconfiguration of the quark substructure of the $\Theta^+$ could be
required for it to decay) are challenged to find a natural physical
mechanism that suppresses the $\Theta^+$ decay more effectively than
the OZI rule suppresses the decay of the $\Lambda(1520)$.

\subsection{Large $N_{\rm c}$ and chiral soliton models}

\subsubsection{Large $N_{\rm c}$}
The number of colors is the only conceivable parameter in light quark
QCD, so it is natural to consider exotic baryon dynamics as an
expansion in $1/N_{\rm c}$.  It is known from the work of `t Hooft,
Witten, and many others, that as $N_{\rm c}\to\infty$ QCD reduces to a
theory of zero width $\bar q q$ mesons with masses $\sim
\Lambda_{QCD}$\cite{thooft} and heavy baryons with masses $\sim
N_{\rm c}\Lambda_{QCD}$ in which quarks move in a mean (Hartree)
field\cite{witten}.  This is far from a complete description even at
the heuristic level: Almost nothing is known about the spectrum of
$\bar q q$ mesons from large $N_{\rm c}$ except for the pseudoscalar
Goldstone bosons required by chiral symmetry.  The dynamics of baryons
is accessible only through a loose association with the chiral soliton
model (CSM).  It is important to remember that the CSM has not be
derived from large $N_{\rm c}$ QCD. Its appeal is based on its proper
implementation of chiral symmetry and anomalies, and on the
\emph{resemblance} of collectively quantized solitons to the lightest
positive parity baryons.  The connection certainly fails for the
simplest case of one flavor, where baryons exist and are described by
a mean field theory (they large $N_{\rm c}$ analogues of the $\Delta^{++}
\equiv uuu$, for example), but there are no Goldstone bosons, no
topology, and no chiral solitons.  On the other hand qualitative
connections between the lowest lying states of a collectively
quantized chiral soliton and the simplest baryons made of $N_{\rm c}$
quarks have been established for arbitrary $N_{\rm f}\ (\ne 0)$ and
large $N_{\rm c}$\cite{manoharcqm}.

Dashen, Jenkins, and Manohar\cite{djm} made the application of large
$N_{\rm c}$ ideas to baryons more precise.  They eschew dynamics and focus
instead on a new $SU(2N_{\rm f})_{\rm c}$ symmetry that emerges as
$N_{\rm c}\to\infty$.  The subscript ``$c$'' means contracted and refers
to a modification of the usual Lie algebra because certain commutators
scale to zero as $N_{\rm c}\to\infty$.  Baryons can be organized into
(infinite dimensional) irreducible representations of this symmetry. 
Jenkins and Manohar show that these irreps can be put into
correspondence with the spectrum of the non-relativistic quark model,
and at least for the ground-state multiplet, with the chiral soliton
model.  Known relationships among baryon masses and couplings can be
classified by the order in $1/N_{\rm c}$ at which they are broken.  New
relations can be derived and are generally quite
successful\cite{jmreview}.  Like any symmetry, however, the utility of
large $N_{\rm c}$ \'a la Jenkins and Manohar is limited by ignorance of
reduced matrix elements.

Jenkins and Manohar have extended their work to exotic
baryons\cite{jmexotics}.  They cannot predict the mass of the lightest
exotic --- it depends on the $SU(2N_{\rm f})_{\rm c}$ invariant
dynamics --- but they can enumerate multiplets that have candidates
for the $\Theta^+$.  They select a positive parity representation
which requires a space wavefunction of mixed symmetry for the $N_{\rm
c}+1$ quarks.  In the language of constituent quarks, the mixed
symmetry space wavefunction corresponds to a state in which $N_{\rm
c}$ quarks are in the Hartree ground state and one is excited.  From a
QCD-inspired quark model viewpoint (see below) this seems like an odd
choice for the ground state multiplet.  To support their choice,
Jenkins and Manohar point out that it is natural in models where
quarks interact by Goldstone boson exchange\cite{flavorqm}.  This
representation of $SU(2N_{\rm f})_{\rm c}$ contains an infinite tower
of $SU(3)$-flavor irreps of definite spin and positive parity.  For
$N_{\rm c}=3$ the lightest $SU(3)_{\rm f}$ multiplet in this
representation is the $\mathbf{\overline{10}}$, where the $\Theta^+$
is expected to lie (see Table \ref{sightings}).  It occurs with
$J^{\Pi}=\frac{1}{2}$ or $\frac{3}{2}$.  Next is a $\mathbf {27}$ with
many exotic candidates\footnote{Actually the situation is a little
more complicated: there are two irreps of $SU(2N_{\rm f})_{\rm c}$
allowed with the mixed symmetry space wavefunction.  One has a tower
of states, $\m{\overline{10}}^{\frac{1}{2}^{+}}$,
$\m{27}^{\frac{1}{2}^{+}}$, $\m{27}^{\frac{3}{2}^{+}}$, \ldots.  The
other has a tower, $\m{\overline{10}}^{\frac{3}{2}^{+}}$,
$\m{27}^{\frac{1}{2}^{+}}$, $\m{27}^{\frac{3}{2}^{+}}$,
$\m{27}^{\frac{5}{2}^{+}}$, \ldots.  The two towers need not be
degenerate.  Jenkins and Manohar only discuss the
first\cite{pirjolthx}.}.

There are also two $SU(2N_{\rm f})_{\rm c}$ irreps in which all the
quarks are in the Hartree ground state --- also a natural candidates
for the $q^{N_{\rm c}}\bar q$ ground state.  All these states have
\emph{negative} parity.  These representations have been studied
recently by Pirjol and Schat\cite{ps}.  They consider both the case in
which all quarks in the exotic are light ($qqqq\bar q$) and the case
where the antiquark is heavy, {\it ie\/} $c$ or $b$ ($qqqq\bar
Q$)\cite{wessling}.  For light quarks the towers begins with
non-exotic $SU(3)_{\rm f} \mathbf{1}\oplus\mathbf{8}$ states, followed
by exotic multiplets (including degenerate (as $N_{\rm c}\to\infty$)
$SU(3)_{\rm f}\ \mathbf{\overline{10}}^{\frac{1}{2}^{-}}$ and
$\mathbf{\overline{10}}^{\frac{3}{2}^{-}}$) at higher mass.

This illustrates a general feature of all quark model/QCD treatments
of the exotic baryons, which I will revisit in the discussion of quark
models (Section III.C): the most natural candidate for the ground
state multiplet (all quarks in the lowest Hartree level) has negative
parity, corresponding to the $KN$ $s$-wave.\footnote{Although the
$d$-wave is possible in principle, a $d$-wave $\Theta^{+}$ does not
occur in the lightest quark model multiplet (see section III.D
below).} So these approaches share a fundamental difficulty at the
outset: If the quark configuration is ``natural'', it's hard to
explain why the $\Theta^+$ should be narrow.  To obtain a narrow
$\Theta^+$, strong quark forces must make a state of mixed spatial
symmetry the lightest.  Understanding these strong quark correlations
and their consequences then becomes a central issue.

The large $N_{\rm c}$ methods of Jenkins and Manohar cannot determine
whether the $\Theta^+$ is light enough to be narrow and prominent, or
even whether it is the lightest $qqqq\bar q$ state.  More dynamical
assumptions are needed for that.  However, if the existence of the
$\Theta^+$ is fed into their machinery, one can predict its properties
and the masses and properties of other states in the $qqqq\bar q$
spectrum\cite{jmexotics}.

\subsubsection{Chiral Soliton Models}

Much of the discussion of exotic baryons over the past
decade\cite{man,chem,bied,pras,weigel} and, in particular, a
remarkable prediction of the mass and width of the $\Theta^+$ by
Diakonov, Petrov, and Polyakov \cite{dpp}, has been carried out in the
context of chiral soliton models.  This is not surprising since CSM's
are teeming with exotics.  Collective quantization of a classical
soliton solution to a chiral field theory of pseudoscalar bosons in
$SU(2)_{\rm f}$ or $SU(3)_{\rm f}$, consistent with anomaly
constraints, yields towers of baryons only the lightest of which
\emph{are not exotic}\cite{anw,guad}.  The simplest example is
$SU(2)_{\rm f}$, where the spectrum of baryons begins with a
rotational band of positive parity states with
$I=J=1/2,3/2,5/2,\ldots$, with masses,
\begin{equation}
	M(I,N_{\rm c})=M_{0}N_{\rm c}+\frac{J(J+1)}{IN_{\rm c}}.
	\label{rigid}
\end{equation}
The parameters $M_{0}$ and $I$ are ${\cal O}(\Lambda_{\rm QCD})$ and
independent of $N_{\rm c}$ as $N_{\rm c}\to\infty$.  Other excitations, radial
for example, are heavier, separated from the ground state band by
${\cal O}(N_{\rm c}^{0})$.  The lack of evidence for a $I=5/2$ baryon
resonance led most workers to dismiss the heavier states as artifacts
of large $N_{\rm c}$.  

The generalization of the CSM to three flavors with broken $SU(3)_{\rm
f}$ has always been controversial.  Guadagnini's original approach
(the ``rigid rotor'' (RR) approach) was to quantize in the $SU(3)_{\rm
f}$ limit and introduce $SU(3)_{\rm f}$ violation
perturbatively\cite{guad}.  Alternatively, Callan and Klebanov
quantized the $SU(2)_{\rm f}$ soliton and constructed strange baryons
as kaon bound states (the ``bound state (BS) approach)\cite{ck}. 
Although different in principle, the two approaches give roughly the
same spectrum for the octet and decuplet.  When generalized to three
flavors the rotational band of the RR approach
becomes\cite{man,chem,bied},
\begin{equation}
 \mathbf{R }^{J^{\Pi}} = \mathbf{8}^{\frac{1}{2}^{+}}, \
 \mathbf{10}^{\frac{3}{2}^{+}}, \
 \mathbf{\overline{10}}^{\frac{1}{2}^{+}}, \
 \mathbf{{27}}^{\frac{1}{2}^{+},\frac{3}{2}^{+}},\ldots
 \label{csmspectrum}
\end{equation}
where $\mathbf{R}$ is the $SU(3)_{\rm f}$ representation.  

Diakonov, Petrov, and Polyakov\cite{dpp} took the first exotic
multiplet in this tower, $\mathbf{\overline{10}}^{1/2^{+}}$,
seriously.  They estimated its mass and width and found that it should
be light and narrow\cite{thetawidth}.  Their work stimulated the
experimenters who found the first evidence for the
$\Theta^+$\cite{spring8}.  Soon after the initial paper by Diakonov
{\it et al} , Weigel examined the spectrum of exotics in the three
flavor CSM more closely\cite{weigel}.  He showed that it is
inconsistent in the RR approach to ignore the mixing between the
$\m{\overline{10}}$ and radial excitations excited by ${\cal O}(N_{\rm
c}^{0})$ above the ground state.  After the first reports of the
$\Theta^+$, other groups looked even more closely at the pedigree of
the $\mathbf{\overline{10}}$ in the CSM\cite{cohen,ikor}.  As Weigel's
analysis had implied, they found the mass splitting between the ground
state and the $\m{\overline{10}}$ to be ${\cal O}(N_{\rm c}^{0})$. 
Cohen pointed out that the width of the $\Theta^+$ does not vanish as
$N_{\rm c}\to \infty$ in contrast to non-exotic states like the
$\Delta$\cite{cohen}, making it hard to understand the very small
value of $\Gamma(\Theta)/\Gamma(\Delta)$ obtained in Ref.~\cite{dpp}. 
Already in 1998 Weigel had pointed out that the $\Theta^+$ does not
exist in the BS approach unless the mass of the kaon is of the order
of 1 GeV\cite{weigel}, a result confirmed by Itzhaki {\it et al\/},
who go on to show that the force between the collectively quantized
two-flavor soliton and the kaon is repulsive for physical kaon masses. 
So either the RR and BS approaches are inconsistent with one another
for the $\m{\overline{10}}$, or the $\m{\overline{10}}$ cannot be
included in the ground state rotational band as in Ref.~\cite{dpp}.
 
Chiral soliton models describe at best a piece of QCD: Their picture
of the nucleon and $\Delta$ (or $\mathbf{8}^{\frac{1}{2}^{+}}$ and
$\mathbf{10}^{\frac{3}{2}^{+}}$ in $SU(3)_{\rm f}$) is internally
consistent and predictive.  Some progress has been made in the
description of baryon resonances\cite{mattiskarliner}.  However the
incorporation of strangeness is not satisfactory and still
controversial\cite{ikor}, and the CSM gives no insight at all into the
meson spectrum.  As for exotics, the candidate for the $\Theta^+$ is
controversial and there is no insight into the striking \emph{absence}
of exotic mesons and baryons in general.  The prediction of a narrow
width for the $\Theta^{+}$ is very controversial.

\subsection{Quark models}

The quark model in its many variations has been by far the most
successful tool for the classification and interpretation of light
hadrons.  It predicts the principal features and many of the
subtleties of the spectrum of both mesons and baryons, and it matches
naturally onto the partonic description of deep inelastic phenomena. 
Perhaps it receives less recognition than it ought to because it
predates QCD. Largely developed during the 1960's by Dalitz and his
students, the quark model was already in place when QCD came on the
scene to legitimized it.  

The limitations of the quark model are, however, as obvious as its
successes.  It has never been formulated in a way that is fully
consistent with confinement and relativity.  Of course quarks can move
relativistically, governed by the Dirac equation, in first quantized
models like the MIT bag\cite{cjjtw}, but there is no fully
relativistic, second quantized version of the quark model. 
Furthermore, quark models are not the first term in a systematic
expansion.  No one knows how to improve on them.

Nevertheless all hadrons can be classified as relatively simple
configurations of a few confined quarks, and there is no reason to be
expect the $\Theta^+$ to be an exception.  So looking for a natural
quark description of the $\Theta^+$ is a high priority, and if there
were none, it would be most surprising.

\subsubsection{Generic features of an uncorrelated quark model}

Although quark models (non-relativistic, bag, flux tube, \ldots)
differ in their details, the qualitative aspects of their spectra are
determined by features that they share in common.  These important
ingredients can be abstracted from the specific models and used to
project expectations for a new sector like $\qqq$.  They need not be
correct --- probably they cannot explain the $\Theta^+$ --- but they
form the context in which other proposals have to be considered. 
Certainly they do a good job for mesons, baryons, and even tetraquark
({\it ie.\/} $\bar q \bar q q q$) spectroscopy.  Here is a summary of
the basic ingredients\cite{jjr,glue}, none of which can be ``derived''
from QCD, with a few words of explanation:
\begin{enumerate}
	\item  The spectrum can be decomposed into sectors in which the 
	numbers of quarks and antiquarks, $n_{q}$ and $n_{\bar q}$, are good
	quantum numbers --- the OZI rule\cite{OZI}.
	
	\item Hadrons are made by filling quark and antiquark orbitals in
	a hypothetical mean field --- a non-relativistic potential or a
	confining bag, for example.  For baryons this might be the Hartree
	mean field suggested by large $N_{\rm c}$, for mesons its origins are
	less clear.  In dual superconductor versions of confinement, like
	the MIT bag model, it is the normal region where colored fields
	are confined.
	
	\item The ground state multiplet is constructed by putting all the
	quarks and antiquarks in the lowest orbital --- the ``single mode
	configuration''.  A natural assumption, but one which must fail
	for $\qqq$ if the $\Theta^{+}$ has positive parity (see
	below).
	
	\item The total angular momentum of the (relativistic) quark in
	the lowest orbital is 1/2.  Its parity is \emph{even} (relative to
	the proton).  Again a natural assumption.  Remember, spin and
	orbital angular momentum are not separately conserved.  Typically
	the first excited orbitals have \emph{negative} parity and total
	angular momentum 1/2 and 3/2 because orbital excitations are
	invariably less costly than radial.
	
	\item The lightest multiplet in any sector, $\bar q^{n_{\bar q}}
	q^{n_{q}}$, can be classified using an $SU(6)$ symmetry built from
	flavor $SU(3)$ and the $SU(2)$ generated by the unitary
	transformations of the $j_{z}=\pm 1/2$ eigenstates connecting the
	lowest $j=1/2$ quark mode.  This generalizes the old $SU(6)$ of
	flavor$\times$spin to relativistic quarks.  For economy of
	notation I will refer to this $SU(2)$ symmetry as ``spin'' and the
	$SU(6)_{\rm fs}$ as ``flavorspin''.  However it is not spin and it
	cannot be used for excited states which include both $j=3/2$ and
	1/2 orbitals without further work\cite{jdg}.
	
	\item These ideas can be applied to the analysis of local, gauge
	invariant operators.  Suppose ${\cal O}$ is a such an operator,
	built from quark and gluon fields.  ${\cal O}$ can create states
	of various spins and parities from the vacuum.  Generically, the
	lower the dimension of ${\cal O}$ the lighter the states it
	creates \cite{jjr} .  If the operator vanishes in the ``single
	mode configuration'', then the states it creates are heavier than
	those created by an operator of the same dimension that does not
	vanish.  Here an example will help: The dimension three operators,
	$\bar q q$, $\bar q\gamma_{5}q$, $\bar q\gamma_{\mu}q$, $\bar
	q\gamma_{\mu}\gamma_{5}q$, and $\bar q\sigma_{\mu\nu}q$, can
	create mesons with $J^{\Pi C}=0^{++},\ 0^{-+},\ 1^{--},\ 1^{+-}$,
	and $1^{++}$.  When the quark fields are replaced by the lowest
	mode, the $0^{++}$, $1^{++}$, and $1^{+-}$ operators vanish,
	leaving $0^{-+}$ and $1^{--}$, which are indeed the lightest meson
	quantum numbers\cite{jjr}.
	 
\end{enumerate}

These are the ingredients in an uncorrelated quark model.  Even though
the quarks can be relativistic, the classification of states and
operators proceeds as if they were non-relativistic.

\subsubsection{Quark model ``states'' and scattering}

As the number of quarks and antiquarks grows, the number of $\bar
q^{n_{\bar q}}q^{n_{q}}$ eigenstates proliferates wildly.  Even in the
ground state multiplet (see 3.  above) there are 36 $\bar q q$ states,
56 $qqq$ states, 666 $\bar q\bar q qq$ states and 1260 $\qqq$ states
(counting each flavor and spin state separately).  The $\bar q q$ and
$qqq$ states are candidates for the lightest mesons and baryons. 
Although the $\bar q\bar q q q$ and $\qqq$ states are stationary
states in a potential or bag, they do not in general correspond to
stable hadrons or even resonances.  Far from it, most, perhaps even
all of them fall apart into $\bar q q$ mesons and $qqq$ baryons
without leaving more than a ripple on the meson-meson or meson-baryon
scattering amplitude.  A $\bar q\bar q qq$ state has the same quantum
numbers and the same quark content as a $\bar q q$-$\bar q q$ meson
scattering state.  In a fairly precise way the $\bar q\bar q q q$
state can be considered a piece of the meson-meson continuum that has
been artificially confined by a confining boundary condition or
potential that is inappropriate in the meson-meson
channel\cite{jl,jl2}.  If the multiquark state is unusually light or
sequestered (by the spin, color and/or flavor structure of the
wavefunction) from the scattering channel, it may be prominent.  If
not, it is just an artifact of a silly way of enumerating the states
in the continuum.
	 
\subsubsection{Pentaquarks in the uncorrelated quark model}

\begin{figure}
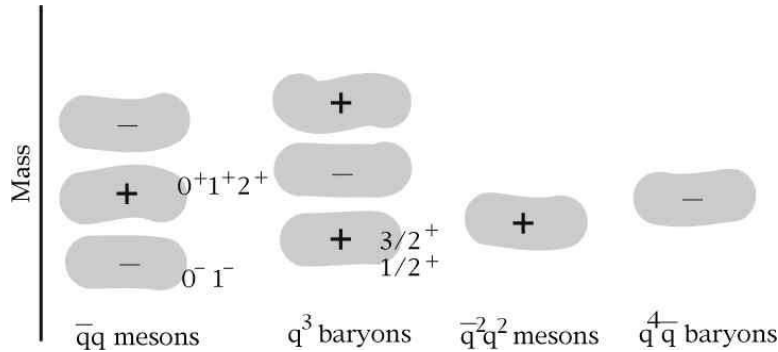

\begin{center}
\BoxedEPSF{quarkparity-.epsf scaled 400} \caption{Pattern of $\bar q q$
meson and $qqq$ baryon parities in the quark model and in Nature.  The
shaded areas represent broad bands of states, the $\pm$ signs label
parity.  When looked at more closely, the bands overlap, but the
general pattern is as shown.  Uncorrelated quark model predictions for
the parity of the lightest $\bar q^{2}q^{2}$ ``tetraquarks'' and
$qqqq\bar q$ pentaquarks are shown schematically.  In other quark
models\cite{flavorqm,jw,kl} dynamics may alter the pattern.}
\label{parities}
\end{center}
\end{figure}

The spectrum of an uncorrelated quark model begins with all quarks and
antiquarks in the same orbital.  Next, one quark or antiquark is
excited, and so forth.  At zeroth order, the spectrum consists of
families of states of alternating parities as shown schematically in
Fig.~\ref{parities}.
The parity of the ground state of $n_{q}$ quarks and $n_{\bar q}$
antiquarks is $(-1)^{n_{\bar q}}$.  When a quark or antiquark is
excited, the parity flips.  Light meson and baryon multiplets
\emph{do} (roughly) alternate in parity --- one of the remarkably
simple and successful predictions of the quark model.  The
pseudoscalar and vector ($\bar q q$) mesons (negative parity) are
followed by $J^{PC}=0^{++}$, $1^{++}$, $1^{+-}$, and $2^{++}$
multiplets; the nucleon octet and decuplet ($qqq$) baryons are
followed by many negative parity multiplets.  There are a few famous
exceptions: for example, the ``Roper'' resonance, with
$J^{\Pi}=1/2^{+}$, is the lightest excited nucleon and the $0^{++}\
f_{0}(600)$ is the lightest excited meson.  (Interestingly, both these
exceptional states are candidates for multiquark states: $\qqq$
for the Roper and $\bar q\bar q qq$ for the $f_{0}(600)$.)  But it is
broadly successful, and so far, it has always got the parity of the
ground state multiplets right.

I will return to the quark model predictions for the lightest
tetraquark states later.  The uncorrelated quark model predicts that
the pentaquark ground state has \emph{negative} parity.\footnote{The
``flavor exchange'' quark model which emphasizes quark-quark forces
mediated by pseudoscalar meson exchange is an exception, one example
of a \emph{correlated} quark model\cite{riskastancu}.  It agrees with
generic quark model predictions of the parity of $\bar q q$ and $qqq$
states, but appears to prefer \emph{positive parity} for the lightest
$\qqq$ states.} This makes the existence of the $\Theta^+$
embarrassing for this model for many reasons:
\begin{enumerate}
	
	\item The $\mathbf{\overline {10}}$ is characteristically
	accompanied by a nearby $\mathbf{8}$ in quark models (see later). 
	The $\mathbf{\overline{10}}$ and $\mathbf{8}$ mix to produce a
	non-strange $\Theta^+$ analogue, $uudd(\bar u, \bar d)$ which should
	be lighter.  There is no candidate for a \emph{negative parity} 
nucleon
	resonance below the $\Theta^+$.
	
	\item In the $\bar q q$ (36 states of spin and flavor) and $qqq$
	(56 states) sectors the ground state multiplets are complete.  In
	the $qqqq\bar q$ sector the ground state multiplet contains 1260
	states.  The $\Theta^+$ and its $SU(3)_{\rm f}$ brethren account only
	for a few --- 36 to be precise.  Of course most will be heavy
	enough to disappear into the continuum.  Still, there are many
	states, and it is hard to imagine that only the antidecuplet
	should be seen.
	
	The counting of states is relatively simple.  Going through it
	will be useful for later purposes: the $q^{4}$ configuration is
	symmetric in space and therefore antisymmetric in
	color$\times$flavor$\times$spin.\footnote{Remember, by ``spin'' I
	really mean the $\uparrow$ and $\downarrow$ states of total
	angular momentum of a $j=1/2$ state.} In color the $q^{4}$ must
	couple to the $\mathbf{3}_{\rm c}$ in order make a singlet with the
	antiquark.  The Young diagram for $[q^{4}]^{\mathbf{3}_{\rm c}}$ is
	shown in Fig.~\ref{q4color} along with the (conjugate) diagram
	that determines the $SU(6)_{\rm fs}$ flavor$\times$spin state.
	\begin{figure}
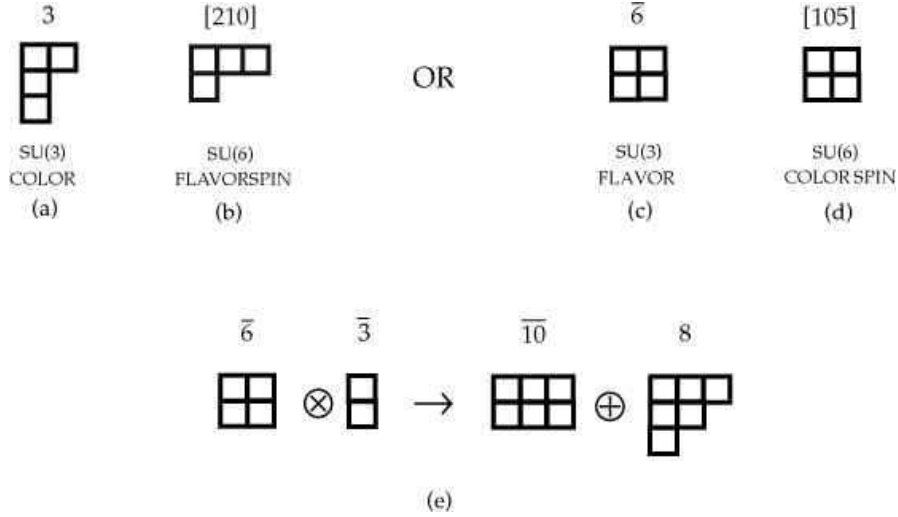

	\begin{center}
\BoxedEPSF{q4color-.epsf scaled 600} \caption{Young diagrams for quark
model exotics.  (a) and (b) show all the negative parity $qqqq\bar q$
states in the space symmetric uncorrelated quark ground state.  The
color must be $\m{3}$ to make a singlet with an antiquark.  The
flavor-spin $SU(6)$ representation that goes with the color $\m{3}$ is
the $[210]$.  It contains only one flavor $SU(3)$ $\m{\overline{6}}$
which has spin 1.  (c) and (d) shows the flavor (c) and $SU(6)$
color-spin (d) diagrams when the $q^4$ flavor is $\m{\overline{6}}$. 
As shown in (e), the $\m{\overline{6}}$ combines with an antiquark,
$\m{\overline{3}}$ to make an exotic $\m{\overline{10}}$ (and a
non-exotic $\m{8}$).  The color spin representation (d) that goes with
the flavor $\m{\overline{6}}$ is the $[105]$.  It contains only one
color $\m{3}$, which has spin 1.  So either way you look at it, there
is a unique operator that creates the negative parity
$\m{\overline{10}}$ with all quarks in the same mode.}
	\label{q4color}
	\end{center}
	\end{figure}

	The
	flavor-spin representation of Fig.~\ref{q4color}(b) is
	210-dimensional and can be decomposed into $SU(3)_{\rm f}\otimes
	SU(2)_{s}$ multiplets (labelled by the dimension of the
	$SU(3)_{\rm f}$ representation and their spin$^{\rm parity}$) as
	follows:
	\begin{equation}
		 [\m{210}]_{\rm fs}= \{\m{3},0^{+}\oplus 1^{+}\} \oplus
		 \{\m{\overline{6}},1^{+}\}\oplus \{\m{15},0^{+}\oplus
		 1^{+}\oplus 2^{+}\}\oplus \{\m{15'},1^{+}\}
		\label{uncorrelated}
	\end{equation}
	These, in turn, must be combined with the antiquark in the
	($\m{\overline{3}}, 1/2^{-}$) representation, and yield an array
	of both exotic and non-exotic negative parity baryons in
	$SU(3)_{\rm f}$ representations including
	$\mathbf{1,8,10,\overline{10},27}$ and $\mathbf{35}$.
	Note that the all-important $\mathbf{\overline{10}}$ comes from 
	\begin{equation}
		 (\m{\overline{6}}, 1^{+})\otimes(\m{\overline{3}},\
		1/2^{-})=(\m{\overline{10}}\oplus
		\m{8},1/2^{-}\oplus 3/2^{-}).  
		\label{quarktenbar}
	\end{equation}
	as illustrated in Fig.~\ref{q4color}(e).  Not only is the
	antidecuplet accompanied by an octet (though their approximate
	degeneracy will not be explained until Section V.D.1), but also
	the candidate for the $\Theta^+$ occurs with both
	$J^{\Pi}=1/2^{-}$ and $3/2^{-}$.  This ``spin-orbit doubling''
	will turn out to be a robust, and problematic, prediction of quark
	models in general\cite{closespinorbit}.
	
	\item A negative parity $\Theta^+$ would have appear in the
	$s$-wave of $KN$ scattering.  Odd parity requires
	$\ell=0,2,4,\ldots$.  The $1/2^{-}$ could only couple to the
	$s$-wave.  Therefore the most promising uncorrelated quark model
	candidate for the $\Theta^+$ would have to appear in the $KN$
	$s$-wave, where its narrow width would be very hard to accomodate.
	
	\item The problem of the width of a negative parity
	$\m{\overline{10}}$ is even more severe than it seems.  However, I
	would like to defer this discussion to the next section, where
	lattice calculations are discussed.
\end{enumerate}

All in all, the uncorrelated quark model gives little reason to expect
a light, narrow, exotic baryon with the quantum numbers of the
$\Theta^+$.  Later I will describe the somewhat more positive
perspective of \emph{correlated} quark models.

\subsection{Early lattice results}

Exotic baryons present an excellent target for lattice QCD: the
existence of the $\Theta^+$ remains in doubt, and its spin and parity
are unknown.  Modern lattice calculations should be able to estimate
the mass of the lightest state in various $\qqq$ channels,
shedding light on the reality of the $\Theta^+$ and its quantum 
numbers,
and predicting other exotics, if they exist.

The first lattice studies of the $\Theta^+$ appeared soon after the
first experimental reports\cite{sazaki,fodor}.  After some initial
confusion about the parity, both groups agreed that there is evidence
for the $KN$ threshold and for a state in the $uudd\bar s$
$J^{\Pi}=1/2^{-}$ channel.  They also reported evidence for a state in
the $1/2^{+}$ channel, but at a higher mass.  Subsequent work by the
Kentucky group\cite{liu} does not find a state in either parity
channel.\footnote{There is another published study\cite{taiwan}, which
claims a $1/2^{+}$ state and general agreement with the diquark model
of Ref.~\cite{jw}, but this work has been criticized for several
technical reasons\cite{jn}.}

These results are troubling: lattice calculations have, in the past,
got the quantum numbers of the ground state correct in each sector of
QCD. In this case the calculations support negative parity
(Ref.~\cite{liu} excepted), which, as we have seen, is hard to
reconcile with the narrow width of the $\Theta^+$.  Several comments 
are
in order:

\begin{enumerate}
	\item Both Refs.~\cite{sazaki} and \cite{fodor} use single, local
	$\qqq$ sources.  For the negative parity
	$\m{\overline{10}}$ channel the local source is unique in a
	certain sense(see below), however for positive parity there are
	eight local sources\cite{jahn} and it is far from clear that the
	chosen source optimizes the overlap with a possible positive
	parity state.  Calculations of the full correlation matrix with
	the eight local sources are underway\cite{jn2}.
	
	\item There is reason to believe, on the basis of diquark ideas,
	that the better positive parity sources may contain explicit
	derivatives, making it ``non-local'' in lattice parlance.  An
	example of such an operator can be found in eq.~(6) of
	Ref.~\cite{jw2}.
	
	\item The calculations are done rather far from the chiral limit. 
	Chiral symmetry is at the heart of the chiral soliton model, where
	the light $1/2^{+}$ $\Theta^+$ was first predicted.  It is also
	important in diquark models, because diquark correlations
	disappear as quark masses increase.  It is not clear whether the
	lattice calculations are close enough to the chiral limit to
	capture the effects that bind the $\Theta^+$.  The $N$-$\Delta$ mass
	difference, which should be a good measure of the importance of
	diquark correlations, is already quite well developed at the quark
	masses used in Refs.~\cite{sazaki} and \cite{fodor}, so perhaps
	they are trustworthy on this score.
\end{enumerate}

Lattice theorists classify sources according to their properties in
the ``non-relativistic'' limit, in which four component Dirac quark
fields are replaced by two component Pauli fields.  This sounds like a
rather uninteresting limit for nearly massless quarks.  However it
applies just as well to the more general quark model outlined in
III.C.1 where the quarks can be massless.  So it can shed light on the
nature of the negative parity $\Theta^+$ that seems to be cropping up 
in
lattice calculations.  In the ``single mode configuration'' the
classification of operators used as lattice sources reduces to the
problem we solved in the previous section for the uncorrelated quark
model.  The important result, summarized by eq.~(\ref{quarktenbar}),
is that there is a \emph{single} $\mathbf{\overline {10}}$ with either
$J^{\Pi}=1/2^{-}$ or $3/2^{-}$.  [In comparison, it is easy to work
out that there are \emph{four} octets with $J^{\Pi}=3/2^{+}$.]

Now to the point: We already know a very simple operator with
$qqqq\bar q$ content and the quantum numbers $(\m{\overline{10}},
1/2^{-})$, namely the local operator that creates a kaon and a nucleon
in the $s$-wave!  No matter what they may look like --- and because of
the complexities of Fierz transformations, they can look completely
different --- any operator that creates a $1/2^{-}$ $\Theta^+$ and
survives the ``single mode approximation'' is just a kaon and an
nucleon on top of one another.  This analysis applies to the operators
used in Refs.~\cite{sazaki,fodor,liu} to create the negative parity
$\Theta^+$.  There is no way to decouple a \emph{negative parity}
$\Theta^+$ source from the $KN$ channel by adroitly superposing
superficially distinct operators, unless one abandons the single mode
approximation, which would presumably add considerable energy to the
state.  On this basis one would expect that the negative parity state
observed in Refs.~\cite{sazaki,fodor,liu} will turn out to have an
enormous width\cite{jahn2}.  The same argument has implications for
the coupling of a negative parity $\Theta^+$ constructed in quark
models: the state has exactly the same spin, color, and flavor
wavefunction as $KN$ in the $s$-wave and therefore should be very
broad.

In summary: lattice studies of $qqqq\bar q$ are only beginning. 
Initial results seem to add to the confusion surrounding the
$\Theta^+$.  They suggest negative parity, but once again the narrow
width of the $\Theta^+$ looks like a major problem.  Advocates of a
positive parity $\Theta^+$ can hope that better sources and better
approximations to the chiral limit will reverse the order of states.

\section{Diquarks}
\setcounter{equation}{0} 

It seems that an uncorrelated quark model leads to a negative parity
ground state multiplet, which contains $1/2^{-}$ and $3/2^{-}$
candidates for the $\Theta^+$.  However the very narrow width of the
$\Theta^+$ seems to be an insuperable difficulty.  So a quark
description of the $\Theta^+$ must look to some correlation to invert
the naive ordering of parity supermultiplets.  This is where diquarks
enter\cite{jw,sn}.

The rest of this paper is devoted to diquarks and their role in
understanding exotics in QCD. As mentioned in the Introduction,
diquarks are not new.  They are almost as old as
QCD\cite{firstdiquarks} and have been the subject of intense study by
many theorists.  The 1993 review by Anselmino {\it et al.\/} gives
references to earlier work\cite{diquarks}.  Their roles in baryon
spectroscopy, in deep inelastic structure
functions, and in dynamics at the confinement
scale\cite{roberts} have been especially emphasized.  Early work on
multiquark states in QCD hinted at the importance of diquarks in
suppressing exotics\cite{rjmulti}, but their wide-ranging importance
in the study of exotics and the {\it aufbau\/} principle of QCD does
not seem to have been recognized previously.

Diquark correlations in hadrons suggest qualitative explanations for
many of the puzzles of exotic hadron spectroscopy: first and foremost,
why exotics are so rare in QCD; next, why the most striking
supernumerary hadrons are a nonet of scalar mesons; third, why an
exotic baryon antidecuplet would be the only prominent baryon exotic;
fourth, why non-strange systems of 6, 9, 12, \ldots quarks form nuclei
not single hadrons; and finally why the $H$ dibaryon ($uuddss$) might
not be as bound as simple estimates suggest.  
% A note of caution,
% however: like all quark model ideas, this one lacks a convincing
% quantitative foundation.  Some features may be confirmed by lattice
% QCD studies in the not-too-distant future, but the need for a
% systematic and predictive phenomenological framework for quark
% spectroscopy has never been greater.

\subsection{Introducing diquarks}

QCD phenomena are dominated by two well known quark correlations:
confinement and chiral symmetry breaking.  Confinement hardly need be
mentioned: color forces only allow quarks and antiquarks correlated
into color singlets.  Chiral symmetry breaking can be viewed as the
consequence of a very strong quark-antiquark correlation in the color,
spin, and flavor singlet channel: $[\bar q q]^{\m{1}_{\rm c}\m{1}_{\rm f} 
0}$. 
The attractive forces in this channel are so strong that $[\bar q
q]^{\m{1}_{\rm c}\m{1}_{\rm f} 0}$ condenses in the vacuum, breaking
$SU(N_{\rm f})_{L}\times SU(N_{\rm f})_{R}$ chiral symmetry.

The ``next most attractive channel'' in QCD seems to be the color
antitriplet, flavor antisymmetric (which is the $\m{\overline{3}}_{\rm
f}$ for three light flavors), spin singlet with even parity:
$[qq]^{\m{\overline{3}}_{\rm c}\m{\overline{3}}_{\rm f} 0^{+}}$.  This
channel is favored by one gluon exchange\cite{dgg,djjk} and by
instanton interactions\cite{thooftinst,shuryak}.  It will play the
central role in the exotic drama to follow.

The classification of diquarks is not entirely trivial.  Several of 
the ideas introduced in Section III.C help us determine which diquark 
configurations are likely to be most attractive and therefore most 
important spectroscopically.

Operators that will create a diquark of any (integer) spin and parity
can be constructed from two quark fields and insertions of the
covariant derivative.  We are interested in potentially low energy
configurations, so we omit the derivatives.  There are eight distinct
diquark multiplets (in color$\times$flavor$\times$spin) that can be
created from the vacuum by operators bilinear in the quark field,
which can be enumerated as follows.  Since each quark is a color
triplet, the pair can form a color $\m{\overline{3}}_{\rm c}$, which
is antisymmetric, or $\m{6}_{\rm c}$, which is symmetric.  The same is
true in $SU(3)$-flavor.  The spin couplings are more complicated. 
Consider $q_{\alpha}q_{\beta}$, where $\alpha$ and $\beta$ are Dirac
indices\cite{rgr}.  The constructions look more familiar if we
represent one of the quarks by the charge conjugate field:
$q_{\alpha}q_{\beta} \to \overline q_{C \alpha}q_{\beta}$, where
$\overline q_{\rm C}=-iq^{T}\sigma^{2}\gamma_{5}$.  Then the
classification of diquark bilinears is analogous to the classification
of $\bar q q$ bilinears: It is easy to show (remembering that the
Dirac fields are in the $(0,1/2)\oplus(1/2,0)$ representation of the
Lorentz group) that $\overline q_{C \alpha}q_{\beta}$ can have spin
zero and one with either even or odd parity: $0^{\pm}$ and $1^{\pm}$.. 
For example, $\bar q_{\rm C}\gamma_{5} q$ creates $0^{+}$ and $\bar
q_{\rm C}\vec\gamma \gamma_{5}q$ creates $1^{-}$.  The parity is
opposite from the more familiar classification of Dirac currents
composed of quark and antiquark.  Furthermore $0^{+}$ and $1^{-}$ even
under quark exchange and $0^{-}$ and $1^{+}$ are odd under quark
exchange.
\begin{table}[htdp] 
\caption{Properties of diquarks.}
\begin{center}
\begin{tabular}{|c|c|c|c| }
\hline
 Spins and Parities & (Flavor, Color) & Operators & Single mode 
survival  \\
\hline
$0^{+}$&$\matrix{\phantom{3^{3^{3}}}(\m{\overline{3}},
\m{\overline{3}})\phantom{3^{3^{3}}}\cr
\phantom{3^{3^{3}}} (\m{6},\m{6})\phantom{3^{3^{3}}}}$& $\overline
q_{\rm C}\gamma_{5}q, \ {\overline q_{\rm C}}\gamma^{0}\gamma_{5}q$&
yes\\
\hline
$1^{+}$&$\matrix{\phantom{3^{3^{3}}}(\m{\overline{3}},
\m{6})\phantom{3^{3^{3}}}\cr\phantom{3^{3^{3}}}
(\m{6},\m{\overline{3}})\phantom{3^{3^{3}}}}$& $\overline q_{\rm
C}\vec\gamma q, \ \overline q_{\rm C}\sigma^{0i} q$& yes\\
\hline
$0^{-}$& $\matrix{
\phantom{3^{3^{3}}}(\m{\overline{3}},\m{6})
\phantom{3^{3^{3}}}\cr\phantom{3^{3^{3}}}
(\m{6},\m{\overline{3}})\phantom{3^{3^{3}}}}$& $\overline q_{\rm C}q,
\ {\overline q_{\rm C}}\gamma^{0}q$& no \\
\hline
$1^{-}$&$\matrix{\phantom{3^{3^{3}}}(\m{\overline{3}},
\m{\overline{3}})\phantom{3^{3^{3}}}\cr\phantom{3^{3^{3}}}
(\m{6},\m{6})\phantom{3^{3^{3}}}}$& $\overline q_{\rm
C}\vec\gamma\gamma_{5}q, \ {\overline q_{\rm C}}\sigma^{ij} q$& no\\
\hline 
\end{tabular}
\end{center}
\label{dirac}
\end{table}%
The eight candidate diquarks are ${\cal A}\{(\m{\overline{3}}_{\rm c}
\oplus\m{6}_{\rm c}) \otimes (\m{\overline{3}}_{\rm c}\oplus\m{6}_{\rm
c})\otimes (0^{\pm}\oplus 1^{\pm})\}$, where ${\cal A}\{\ldots\}$
denotes the antismmetrization required by fermi statistics.  Their
properties are summarized in Table~\ref{dirac}.  The candidates can be
pared down quickly:
\begin{itemize}
	\item Color $\m{6}_{\rm c}$ diquarks have much larger color
	electrostatic field energy.  All models agree that this is not a 
	favored configuration.  
	
	\item Odd parity diquark operators vanish identically in the
	 single mode configuration.  The reasons for using this
	criterion were described in the previous section: it
	corresponds to quarks that are unexcited relative to one another.  
 
% 
% 	%
% 	\begin{equation}
% 		q(x,t) \to 
% 		 \pmatrix{
% 		if(r)u_{m} \cr
% 		\vec\sigma\cdot \hat r g(r)u_{m}}e^{-i\omega t}
% 	\qquad
% 		q_{\rm c}(x,t) \to
% 	\pmatrix{
% 		\vec\sigma\cdot \hat r g(r)u_{m}\cr -if(r)u_{m} 
% 		}e^{-i\omega t}
% 	\end{equation}
% 	%	
% 	(where $f(r)$ and $g(r)$ are functions of $|\vec x|$ only, and
% 	$u_{m}$ is a two component Pauli spinor) and form the operators
% 	corresponding to the $0^{-}$ or $1^{-}$ diquark.  They will
% 	vanish.  This indicates that to form negative parity the two
% 	quarks must be placed in different modes of a mean field, or
% 	equivalently, given relative excitation.  All models also agree
% 	that negative parity diquarks are not an attractive configuration.
\end{itemize}

This leaves only two diquarks,
\begin{eqnarray}
	&&\left|\{qq\}\ \m{\overline{3}}_{\rm c}(A)\ \m{\overline{3}}_{\rm f} 
(A)\
	0^{+}(A)\right\rangle \nonumber\\
	&&\left|\{qq\}\ \m{\overline{3}}_{\rm c}(A)\ \m{ 6}_{\rm f}(S)\
	1^{+}(S)\right\rangle
\end{eqnarray}
where $A$ or $S$ denotes the exchange symmetry of the preceding
representation.  Both of these configurations are important in
spectroscopy.  In what follows I will refer to them sometimes as the
``scalar'' and ``vector'' diquarks, or more suggestively, as the
``good'' and ``bad'' diquarks.  Remember, though, that there are many
``worse'' diquarks that we are ignoring entirely.  As an example of
the process by which operators are constructed, here is the good
diquark operator,
\begin{equation}
	\qq_{ia}=\epsilon_{ijk}\epsilon_{abc
	}(i\sigma_{2})_{\alpha\beta}\ 
	q^{jb}_{\alpha}q^{kc}_{\beta}=\epsilon_{ijk}\epsilon_{abc}\ \overline
	q^{jb}_{\rm c}\gamma_{5}q^{kc}
	\label{diquarkoperator}
\end{equation}
Models universally suggest that the scalar diquark is lighter than the
vector.  For example, one gluon exchange evaluated in a quark model
gives rise to a color and spin\footnote{Here again ``spin'' refers to
the total angular momentum of a $1/2^{+}$ quark, which coincides with
spin only in the non-relativistic limit.} dependent interaction,
\begin{equation}
	{\cal H}_{\rm color\ spin}= - \alpha_{s}\sum_{i\ne
	j}M_{ij}\vec\sigma_{i}\cdot \vec\sigma_{j}
	\tilde\lambda_{i}\cdot\tilde\lambda_{j}
	\label{colorspin}
\end{equation}
where $\vec\sigma_{i}$ and $\tilde\lambda_{i}$ are the Pauli and
Gell-Mann matrices that operate in the spin and color space of the
$i^{\rm th}$ quark.  $M_{ij}$ is a model-dependent matrix element that
depends on the mass of the quarks but not on their spin and color. 
$M_{ij}$ is largest for massless quarks ($M_{oo}$), decreases if one
($M_{os}$) or both ($M_{ss}$) of the quarks are strange, and decreases
like $1/m_{i}m_{j}$ for heavy quarks.  Ignoring quark mass effects the
matrix elements of this operator in the ``good'' and ``bad'' diquark
states are $-8M_{oo}$ and $+8/3M_{oo}$ respectively.  To set the
scale, the $\Delta$--nucleon mass difference is $16M_{oo}$, so the
energy difference between good and bad diquarks is $\sim \frac{2}{3}
(M_{\Delta}-M_{N})\sim$ 200 MeV. Not a huge effect, but large enough
to make a significant difference in spectroscopy.  After all, the
nucleon is stable and the $\Delta$ is 300 MeV heavier and has a width
of 120 MeV!

\subsection{Characterizing diquarks}

\begin{figure}
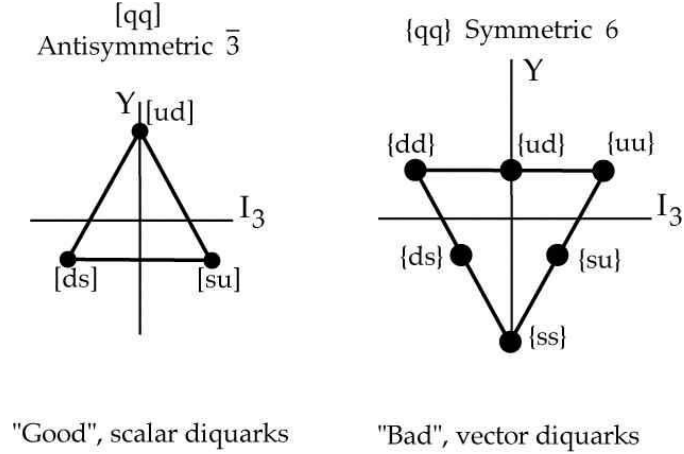

	\begin{center}
	\BoxedEPSF{diquarkqn-.epsf scaled 400} \caption{$SU(3)$ weight
	diagrams and quark content of the ``good'' scalar diquarks and the
	``bad'' vector diquarks.  Braces denote symmetrization and
	brackets denote antisymmetrization in flavor.}
	\label{diquarkqn}
	\end{center}
	\end{figure}

The good scalar and bad vector diquarks are our principal subjects. 
Their quark content and flavor quantum numbers are summarized in
Fig.~\ref{diquarkqn}.  Since the good diquarks are antisymmetric in
flavor, we will denote them by $[q_{1},q_{2}]: \{[u,d]\ [d,s]\
[s,u]\}$ when flavor is important and by $Q\!\!\!\!Q$ when it is not. 
It is particularly easy to construct flavor wavefunctions for baryons,
pentaquarks, {\it etc\/}, made of good diquarks by using the
correspondence
\begin{eqnarray}
	[u,d]& \leftrightarrow &  \bar s\nonumber\\[0ex]
	[d,s]& \leftrightarrow & \bar u\nonumber\\[0ex]
	[s,u]& \leftrightarrow & \bar d
	\label{correspondences}
\end{eqnarray}
which is obvious from Fig.~\ref{diquarkqn}\footnote{Note, however,
that the signs in eqs.~(\ref{correspondences}) are important.  They
are determined by cyclic permutation.  If you use $[u,s]$ instead of
$[s,u]$ you will get into sign trouble!}.  The bad diquarks are
symmetric in flavor, suggesting the notation
$\{q_{1},q_{2}\}:\{\{u,u\} \ \{u,d\}\ \{d,d\}\ \{d,s\}\ \{s,s\} \
\{s,u\}\}$.

% For completeness I also list, in Table~\ref{diracqn} all the diquark
% operators of the form $\bar q_{\rm c}\Gamma q$ and the spin-parity of 
%the
% states that they create from the vacuum.  To see which survive in 
%the
% ``single mode limit'' discussed above, substitute
% eq.~(\ref{singlemode}).  The survivors are noted in the table.
	%

Diquarks are, of course, colored states, and therefore not physical. 
However, their properties can be studied in a formally correct, color
gauge invariant way on the lattice.  To define the non-strange
diquarks, introduce an infinitely heavy quark, $Q$, {\it ie\/} a
Polyakov line.  Then study the $qqQ$ correlator with the $qq$ quarks
either antisymmetric ($ [u,d]Q $) or symmetric ($\{u,d\}Q$) in flavor. 
The results, $M[u,d]$ and $M\{u,d\}$ --- labelled unambiguously ---
are meaningful in comparison, for example, with the mass of the
lightest $\bar q Q$ meson, $M(u)=M(d)$\footnote{I ignore small isospin
violating effects throughout.}.  $ M\{u,d\}- M[u,d]$ is the good-bad
diquark mass difference for massless quarks.  It a is measure of the
strength of the diquark correlation.  The diquark-quark mass
difference, $M[u,d]-M(u)$, is another.  The same analysis can be
applied to diquarks made from one light and one strange quark giving
$M[u,s]$ and $M\{u,s\}$.  The mass of the doubly strange vector
diquark, $M\{s,s\}$ can be measured similarly.  These mass differences
are \emph{fundamental} characteristics of QCD, which should be
measured carefully on the lattice.

Of course, a moment's thought reveals that $M[u,d]$ is the mass of the
particle usually called the $\Lambda_{Q}$.  $M\{u,d\}$ corresponds to
the $\Sigma_{Q}$ ($J=1/2$) or $\Sigma^{*}_{Q}$ ($J=3/2$), the two
being degenerate in the infinite quark mass limit.  $M[u,s]$ and
$M\{u,s\}$ are related to masses of $\Xi_{Q}$s and $\Xi^{*}_{Q}$s; 
and $M\{s,s\}$ to the masses of the $\Omega_{Q}$s\cite{naming}.

Lattice calculations of these quantities may take a while, and will be
subject to debates about how close they are to the chiral and
continuum limits.  In the meantime we already know enough about the
masses of charm mesons and baryons to extract an estimate of these
mass differences, although we are handicapped by the fact that the
spin interactions between the light quarks and the charm quark are not
negligible.  It would be even better if we knew the masses of enough
bottom baryons to perform the analysis with the $b$-quark as the heavy
spectator.  So far only enough is known about $b$-baryon masses to
extract the diquark-quark mass difference.  Of course the scalar
diquark has no spin interaction with the spectator heavy quark ($Q$),
but the vector diquark does,
\begin{equation}
	{\cal H}(Q,\{q_{1},q_{2}\}) = K(Q,\{q_{1},q_{2}\})2\vec
	S_{\{q_{1},q_{2}\}}\cdot \vec S_{Q}
	\label{spinspin}
\end{equation}
where $\vec S_{\{q_{1},q_{2}\}}$ is the spin of the vector diquark,
and the coefficient $K(Q,\{q_{1},q_{2}\})$ depends on the quark
masses.  The light antiquark and heavy quark in a $\bar q Q$ meson 
has a similar
interaction,
\begin{equation}
	{\cal H}(Q,\bar q)=K(Q,\bar q) 	2
\vec s_{\bar q}\cdot \vec S_{Q}
	\label{antispinspin}
\end{equation}
This interaction splits $D^{*}_{Q}$ from the $D_{Q}$,\footnote{I
denote the $(Q\bar u/\bar d)$ pseudoscalar and vector mesons as
$D_{Q}$ and $D^{*}_{Q}$ respectively, and the $(Q\bar s)$ mesons as
$D_{sQ}$ and $D^{*}_{sQ}$.} the $\Sigma^{*}_{Q}$ from the $\Sigma_{Q}$
and the $\Xi^{*}_{Q}$ from the $\Xi_{Q}$ and $\Xi'_{Q}$.  Other spin
dependent interactions mix the $\Xi_{Q}=[[u,s]Q]^{J=1/2}$ with the
$\Xi_{Q}'=[\{u,s\}Q]^{J=1/2}$.

In order to obtain estimates of diquark mass differences, it is
necessary to take linear combinations of baryon and meson masses that
eliminate these spin interactions.  Among the non-strange quarks, we
obtain
\begin{eqnarray}
	M\{u,d\}|_{Q}-M[u,d]|_{Q}&=&\frac{1}{3}\left(
	2M(\Sigma^{*}_{Q})+M(\Sigma_{Q})\right) -M(\Lambda_{Q})\nonumber\\
	M[u,d]|_{Q}- M(u)|_{Q}&=&M(\Lambda_{Q})-
	\frac{1}{4}\left(M(D_{Q})+3M(D^{*}_{Q})\right)	\nonumber\\
	K(Q,\{u,d\}) &=
	&\frac{1}{3}\left(M(\Sigma^{*}_{Q})-M(\Sigma_{Q})\right)
	\label{nonstrange}
\end{eqnarray}

To obtain useful information from the $\Xi_{Q}$ and $\Omega_{Q}$
($\Omega=(Qss)^{J=1/2}$) states, it is necessary to assume that both
the bad diquark mass and the spin interaction are linear functions of
the strange quark mass, 
\begin{eqnarray}
	M\{s,s\}|_{Q}+M\{u,d\}|_{Q}&=&2M\{u,s\}|_{Q}\nonumber\\
	K(Q,\{s,s\})+K(Q,\{u,d\})&=&2K(Q,\{u,s\})\ ,
	\label{linear}
\end{eqnarray}
amounting to first order perturbation theory in $m_{s}$.  With this we
can deduce,
\begin{eqnarray}
 \left.M\{u,s\}\right|_{Q}-\left.M[u,s]
 \right|_{Q}&=&\frac{2}{3}\left( M(\Xi^{*}_{Q})+M(\Sigma_{Q})
 +M(\Omega_{Q} ) \right) -M(\Xi_{Q})- M(\Xi'_{Q})\nonumber\\
	M[u,s]|_{Q}-M(s)|_{Q}&=&M(\Xi_{Q})+M(\Xi'_{Q})
	-\frac{1}{2}(M(\Sigma_{Q})+M(\Omega_{Q}))
	-\frac{1}{4}(M(D_{sQ})+3M(D^{*}_{sQ}))\nonumber\\
	K(Q,\{u,s\}) &= &\frac{1}{6}\left(2M(\Xi^{*}_{Q})-M(\Omega_{Q})
	-M(\Sigma_{Q})\right).
	\label{strangecharm}
\end{eqnarray}

This analysis can be applied directly in the charm sector, where all
the required hadron masses are known.  Only the middle of
eqs.~(\ref{nonstrange}) can be applied in the bottom sector due to
lack of information about bottom baryons.  Finally, if we are daring,
we can apply the first part of this analysis, eqs.~(\ref{nonstrange}),
to the strange baryons.  This is not as ill-founded as it might seem,
since we are not ignoring the spin-spin interactions between the light
($u$ and $d$) quarks and the $s$-quark.  However, there is no useful
analogue of eqs.~(\ref{strangecharm}) in the strange sector --- as the
absence of a strange state with the symmetry structure of the
$\Omega_{Q}$ should make clear --- without making a more detailed
model\footnote{In fact, quark models suggest a more microscopic model
in which all residual quark interactions are described by a spin-spin
interaction, ${\cal H} = \sum_{i\ne j}K_{ij}\vec s_{i}\cdot \vec
s_{j}$.  The reader is invited to work out the diquark masses in this
model.}.
 
When we substitute numbers into 
eqs.~(\ref{nonstrange})--(\ref{strangecharm}), quite a consistent picture 
of 
diquark mass differences and diquark-spectator interactions emerges:
First,
\begin{eqnarray}
	M\{u,d\}|_{s}-M[u,d]|_{s}&=& 205 \ \mbox{MeV}\nonumber\\
	M\{u,d\}|_{\rm c}-M[u,d]|_{\rm c}&=& 212 \ \mbox{MeV}\nonumber\\[3ex]
	M[u,d]|_{s}-M(u)|_{s} &=& 321 \ \mbox{MeV}\nonumber\\
	M[u,d]|_{\rm c}-M(u)|_{\rm c} &=& 312 \ \mbox{MeV}\nonumber\\
		M[u,d]|_{b}-M(u)|_{b} &=& 310 \ \mbox{MeV}
	\label{diquarkmassdiffs}
\end{eqnarray}
shows that the properties of hypothetical non-strange diquarks are the
pretty much the same when extracted from the charm and bottom, and
even strange, baryon sectors.  Second,
\begin{eqnarray} 
	M\{u,s\}|_{\rm c}-M[u,s]|_{\rm c}&=& 152 \ \mbox{MeV}\nonumber\\
	M[u,s]|_{\rm c}-M(s)|_{\rm c} &=& 498 \ \mbox{MeV}
	\label{sdiquarkmassdiffs}
\end{eqnarray}
shows that the diquark correlation decreases when one of the light
quarks is strange.  This is certainly to be expected, since it
originates in spin dependent forces.  As the correlation decreases the
mass difference between the scalar and vector diquarks decreases
($\sim$210$\to\sim$150 MeV) and the mass difference between the scalar
diquark and the antiquark increases ($\sim$310$\to\sim$500 MeV). 
Finally,
\begin{eqnarray}
	K(s,\{u,d\}) &= & 64 \ \mbox{MeV}\nonumber\\
	K(c,\{u,d\}) &= & 21 \ \mbox{MeV}\nonumber\\
	K(c,\{u,s\}) & = & 24 \ \mbox{MeV}
	\label{spinsize}
\end{eqnarray}
shows that the non-strange vector diquark interaction with the
spectator charm quark is significantly weaker than with a spectator
strange quark, as expected from heavy quark theory.  The only mildly
surprising result is that the $\{u,s\}$ and $\{u,d\}$ vector diquarks
have roughly the same interaction with the charm spectator.  It will
be very interesting to compare these results with further measurements
in the $b$-quark sector and, of course, with the results of lattice
calculations.

\subsection{Phenomenological evidence for diquarks}

\begin{figure}
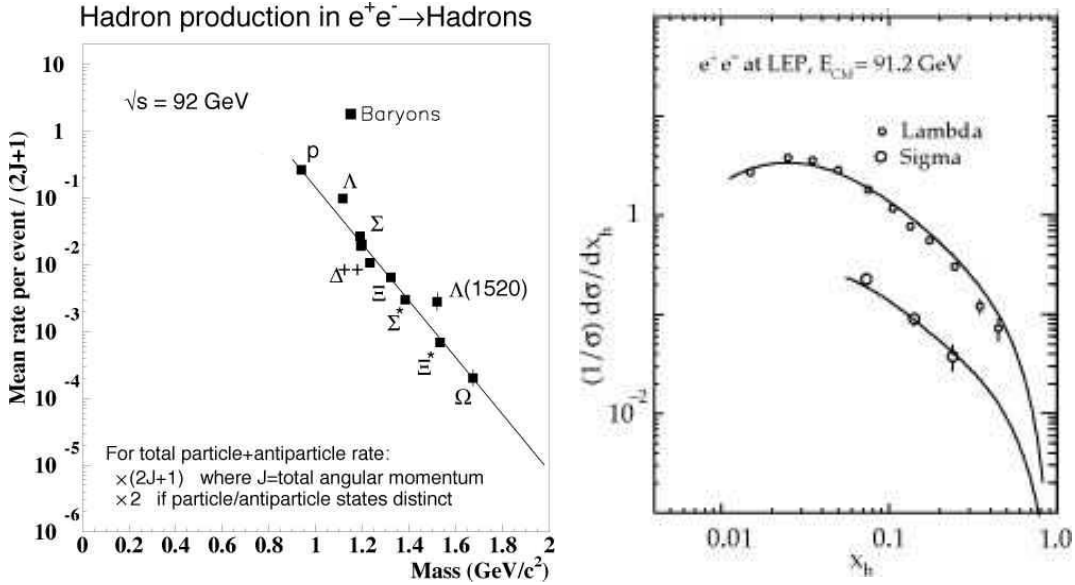

\begin{center}
\BoxedEPSF{lambdafrag-.epsf scaled 400}\quad 
\BoxedEPSF{delphi-.epsf
scaled 840} \caption{(a) Total inclusive baryon production in
$e^{+}e^{-}$ production at LEP energies\cite{thanks}.  The exponential
line is only to guide the eye.  (b) A comparison of the $\Lambda$ and
$\Sigma$ fragmentation functions measured at DELPHI\cite{delphi}.}
\label{lambdafrag}
\end{center}
\end{figure}

Once you start looking, there is evidence for diquarks everywhere.  
What follows is hardly more than a list to whet the appetite, with 
occasional explanations. 
\begin{itemize}
	\item  Baryon spectroscopy.
	
	 Diquarks were born in the regularities of the baryon
	 spectrum\cite{firstdiquarks}, which seem be described
	 qualitatively by viewing baryons as quark-diquark bound states. 
	 In addition to the spin splittings that were described in the
	 previous subsection, another famous piece of evidence is the
	 apparent absence of an $ [\m{20}]_{\rm fs} $ in the second
	 ``band'' of excited baryons in the traditional quark model
	 $O(3)\times SU(6)_{\rm fs}$ classification of
	 baryons\cite{20plet}.  In the zeroth and first bands ( $L=0\
	 [\mathbf{56}]_{\rm fs}$ or $L=1\ [\mathbf{70}]_{\rm fs}$) the
	 $qqq$ space wavefunction is either symmetric or of mixed
	 symmetry, allowing pairs of quarks to form correlated diquarks. 
	 Only in the second band of excited baryons is it possible to have
	 a totally antisymmetric space wavefunction, which cannot be made
	 of a quark-diquark pair.  Since the $qqq$ state is by hypothesis
	 antisymmetric in space and already antisymmetric in color, it
	 must be antisymmetric in flavor $\times$ spin, which is the
	 $[\mathbf{20}]_{\rm fs} $.  The $[\mathbf{20}]_{\rm fs}$ contains
	 a $(\mathbf{8}_{\rm f},\frac{1}{2})$ and a $(\mathbf{1}_{\rm
	 f},\frac{3}{2})$.  This is the only spin-3/2, flavor singlet in
	 any $qqq$ $SU(6)_{\rm fs}$ multiplet.  Its absence is noted in
	 the PDG review \cite{20plet} of the quark model classification of
	 baryon resonances.\footnote{Amsler and Wohl suggest that members
	 of the $[\mathbf{20}]_{\rm fs}$ would be hard to produce ``since
	 a coupling to the ground state would require a two-quark
	 excitation'', although this is not a well tested dynamical
	 principle.}
	 
	 \item The $\Delta I=1/2$ rule in weak non-leptonic decays.
	 
	 The four-quark ($\bar q\Gamma q\ \bar q\Gamma q$) operators in
	 the effective Lagrangian for weak non-leptonic decays transform
	 with either $I=1/2$ or $I=3/2$ (or, in the case of three flavors
	 $\m{8}$ or $\m{27}$).  These operators mediate $K\to 2\pi$, $K\to
	 3\pi$ decays as well as hyperon decays like $\Lambda\to N\pi$,
	 $\Sigma\to N\pi$, {\it etc.\/} The $I=1/2$ operator appears to be
	 enhanced over the $I=3/2$ operator by over an order of magnitude. 
	 A small part of that enhancement can be attributed to the
	 perturbative evolution of the operator from the weak ($M_{Z}$)
	 scale down to the QCD scale\cite{pertevol}.  The rest of the
	 enhancement is presumably non-perturbative in origin.  The
	 $I=1/2$ operator is built from good, scalar diquarks, the $I=3/2$
	 involves bad, vector diquarks.  In the 1980's Neubert, Stech, and
	 their collaborators showed how a systematic enhancement of the
	 scalar diquark over the vector would explain the $\Delta I=1/2$
	 rule in both meson and baryon non-leptonic weak
	 decays\cite{neubert}.
	 
	 \item Regularities in parton distribution functions.
	 
	 The famous 4:1 ratio of proton and neutron deep inelastic 
	 structure functions as $x\to 1$ 
	 \begin{equation}
		 \lim_{x\to 1} \frac{F_{2}^{\rm en}(x,Q^{2})}
		 {F_{2}^{\rm ep}(x,Q^{2})} = \frac{1}{4}
		 \label{1/4rule}
	 \end{equation}
	 follows from the dominance of the scalar diquark\cite{feynman,
	 closethomas}.  Data are not available all the way to $x=1$, but
	 the tendency for the data to decrease toward the positivity bound
	 of 1/4 is clear.  As the Bjorken-$x$ of the struck quark
	 approaches one, the two spectator quarks are forced to their most
	 tightly bound configuration.  If the scalar diquark dominates
	 then only the $u$ quark in the proton and the $d$ quark in the
	 neutron can survive as $x\to 1$.  The 1/4 is the ratio of their
	 squared charges.  Similar regularities are predicted for spin
	 dependent structure functions\cite{closethomas},
	 \begin{eqnarray}
		 \lim_{x\to 1}\frac{\Delta d}{d} &=& -\frac{1}{3}\nonumber\\
		 \lim_{x\to 1}\frac{\Delta u}{u} &=& 1
	 \label{spinrules}
	 \end{eqnarray}
	 New data coming out of low-$Q^{2}$ inelastic electron scattering
	 experiments at JLab seem to support these
	 predictions\cite{jlabxto1}.
	 
	 \item Quark condensation in dense matter.
	 
	 The phenomenon of color superconductivity in dense quark matter
	 has attracted widespread attention over the past few
	 years\cite{colorsuper}.  It does not qualify as phenomenological support
	 for diquarks because no one has figured out how to observe cold
	 quark matter at high baryon density.  However, the fundamental
	 Cooper pair of color superconductivity is the good scalar
	 diquark.  In dense enough matter one can \emph{prove} that this
	 correlated scalar diquark is so tightly bound that it condenses,
	 breaking color$\times$flavor down to a subgroup with many
	 interesting, if hard to observe, consequences.
	 
	 \item $\Lambda(1116)$ and $\Lambda(1520)$ fragmentation
	 functions.

	 The $\Lambda(1116)$ is special among stable baryons.  Because it
	 is an isosinglet, the $ud$ pair is 100\% in the good, scalar
	 diquark configuration.  The $\Sigma(1192)$ resembles the
	 $\Lambda$ except for a less favorable diquark content.  It is
	 therefore interesting to compare their production in a clean
	 environment, like fragmentation in $e^{+}e^{-}$ annihilation at
	 LEP, where baryons are seen as fragments of the quarks produced
	 in $e^{+}e^{-}\to q \bar q$.  A summary of the production cross
	 sections for various particles can be found in the
	 Ref.~\cite{pdg}.  Typically cross sections fall roughly
	 exponentially with the mass of the produced hadron (about a
	 factor of $e$ for every 100 MeV\cite{thanks}).  The
	 $\Lambda(1116)$ is the only exception among stable baryons: it is
	 produced about 2-3 times more copiously than one would expect, as
	 shown in Fig.~\ref{lambdafrag}(a).  Interestingly, the
	 $\Lambda(1520)$ resonance is also anomalously abundant.  This
	 provides even further support for the dominance of the good
	 diquark: The $\Lambda(1520)$ can be described as a good $[u,d]$
	 diquark and an $s$ quark in a $p$-wave\cite{sw}.
	 
	 The contrast between the $\Lambda(1116)$ and the $\Sigma(1192)$
	 is even more striking when one compares the cross sections as a
	 function of $z$, the fractional energy of the baryon.  At large
	 $z$ where the valence component of hadron wavefunctions dominate,
	 the $\Lambda/\Sigma$ ratio, as measured at LEP\cite{delphi} is
	 more than an order of magnitude (see Fig.~\ref{lambdafrag}(b)).
\end{itemize}

\subsection{Diquarks and higher twist}

Diquarks need not be pointlike.  As we have seen, the energy
difference between the good and bad diquarks is only $\sim$ 200 MeV,
enough to be quite important in spectroscopy, but corresponding only
to a correlation length of 1 fermi, the same as every other mass scale
in QCD. It is interesting, nevertheless to ask whether other hadronic
phenomena can constrain the correlation.  Although many nucleon
properties, like form factors, are often discussed in terms of quark
correlations, as far as I know, the correspondence can only be made
exact for deep inelastic scattering (DIS).

Any kind of quasi-pointlike ({\it ie\/} characterized by a mass scale
$\Lambda_{\qq}\gg \Lambda_{\rm QCD}$) correlation in the nucleon is
certainly excluded for $\Lambda_{\qq}$ ranging from $\sim$ 1 GeV up to
the highest scales where deep inelastic data exist ($\sim$100 GeV). 
Diquarks would be especially obvious because as bosons they would
generate an anomalously large longitudinal/transverse inelastic cross
section ratio in DIS at scales below $\Lambda_{\qq}$, which would
disappear above $\Lambda_{\qq}$.  Such an effect is certainly ruled
out by the early, and apparently permanent, onset of scaling seen in a
multitude of experiments.

\begin{figure}
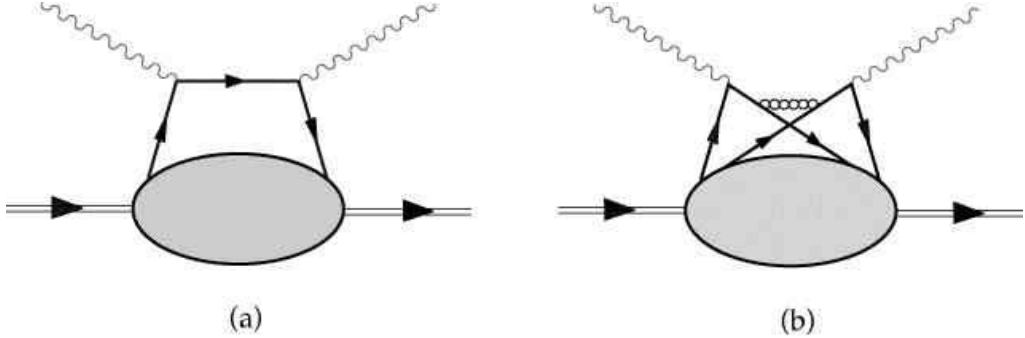

\begin{center}
\BoxedEPSF{dis-.epsf scaled 600} 
\caption{(a) Leading twist, single quark contribution to DIS  (b) 
Twist-4, diquark contribution to DIS.}
\label{dis}
\end{center}
\end{figure}

On the other hand one might think that the absence of large higher
twist effects in DIS could be used to place an uncomfortably
\emph{low} limit on the mass scale of diquark correlations.  This is
not the case\cite{jv}.  In fact measurements of $1/Q^{2}$ corrections
to DIS \emph{place no limits whatsoever} on scalar diquark
correlations in the nucleon.  To understand this it is necessary to
review some of the basics of the twist analysis of deep inelastic
scattering.  ``Twist'' refers to the dimension ($d$) minus the spin
($n$) of the operators that contribute to DIS, $t=d-n$.  The smaller
the twist, the more important the contribution to DIS: A given
operator contributes like $1/Q^{t-2}$.  The leading operators are
twist-2 and act on a single quark\footnote{I am ignoring gluon
operators, which do not figure in the argument.}.  They have the
generic structure
\begin{equation}
	{\cal O}^{(2)}_{\mu}\sim \bar q \gamma_{\mu}{\cal D}{\cal D}\ldots q
	\label{twist2}
\end{equation}
The covariant derivatives, their Lorentz indices suppressed, denoted
schematically by ${\cal D}$, have $d({\cD})-n({\cD})=0$, so they are
irrelevant for counting twist.  The quark fields have $d(q)=3/2$ and
the $\gamma$-matrix contributes $n(\gamma)=1$, so in all,
$t=2(3/2)-1=2$, and these operators' contributions to DIS are
independent of $Q$ (modulo logarithmic corrections from perturbative
QCD).  The $\bar q \gamma q$ operators sum up to give the ``handbag''
diagram shown in Fig.~\ref{dis}(a).

It is easy to write down operators with twist greater than
two\cite{js}.  The most important are twist-four (twist-three does not
contribute to spin average DIS for light quarks), which contribute
corrections of order $1/Q^{2}$ to deep inelastic structure functions. 
The factor of $1/Q^{2}$ is accompanied by some squared mass-scale,
$M_{4}^{2}$, in the numerator.  Twist-four effects have been studied
for years, and the qualitative conclusion is that $M_{4}$ is small. 
How small need not concern us, for we are about to see that it anyway
places no limit on the good diquark that interests us.

Twist four operators invariably involve products of more than two
quark and gluon fields (again ignoring pure-gluon operators). 
Examples include quark-gluon operators, $\bar q {\cal F} q$ and $\bar
q {\cal F}{\cal F} q$, and four-quark operators, $\bar q q \bar q q$. 
The matrix elements of these operators in the target nucleon determine
the magnitude of higher twist effects.  The four quark operators are
the culprits: they can be Fierz-transformed into diquark-antidiquark
operators, $\bar q\bar q \ldots q q$ and therefore measure the scale
of diquark correlations in the nucleon.  They can be summed (in a
well-defined way) to give diagrams like Fig.~\ref{dis}(b), where two
quarks are removed from the nucleon, scattered at high momentum, and
then returned\cite{partonsattwist4}.  The generic structure of four 
quark operators is
(there are others, but the results are the same),
\begin{equation}
	{\cal O}^{(4)}_{\mu\nu}\sim \bar q\gamma_{\mu}{\cal D}{\cal D}\ldots 
	q\ \bar q\gamma_{\nu}{\cal D}{\cal D}\ldots q.
	\label{twist4}
\end{equation}
The $\gamma$-matrices are necessary.  With $d(q)=3/2$ and $d({\cal
D})-n({\cal D})=0$ it is easy to see that the twist of ${\cal
O}^{(4)}$ would be six if it were not for the two factors of $\gamma$,
each of which corresponds to a unit of spin.  In other words: when
Fierzed, the two diquarks in ${\cal O}^{(4)}$ must be coupled to
spin-2.  So \emph{only the vector diquark contributes at twist-four}. 
Bounds on twist four in DIS tell us that the bad, vector diquark
cannot be tightly bound, but they do not constrain the good, scalar
diquark at all.  It contributes only to twist-six and beyond, where it
cannot be separated from the flood of non-perturbative effects that
emerge at low $Q^{2}$.

We can proceed without concern that correlations of the extent
necessary to influence the spectrum are ruled out by deep inelastic
phenomena.

\section{Diquarks and Exotics}
\setcounter{equation}{0} 
\subsection{An overview}

Let us consider exotic spectroscopy with diquark correlations in mind. 
I will assume little more than that two quarks prefer to form the
good, scalar diquark when possible.  States dominated by that
configuration should be systematically lighter, more stable, and
therefore more prominent, than states formed from other types of
diquarks.  This qualitative rule leads to qualitative predictions ---
all of which seem to be supported by the present state of experiment. 
This is clearly an idealization --- a starting place for describing
exotic spectroscopy.  Important effects are ignored, for example
residual QCD interactions can turn a scalar diquark into a vector
diquark\footnote{Because gluons are flavor singlets they cannot
transform a good, $\mathbf{\overline{3}}_{\rm f}$-diquark into a bad,
$\mathbf{6}_{\rm f}$-diquark.  Instead quark exchange between the
diquarks is required as well.}.  A more sophisticated treatment would
have to consider these effects quantitatively.  In fact, the scheme I
am describing here is a step back in complexity --- though it may
capture the underlying physics better --- from the first work on
multiquark spectroscopy in QCD\cite{rjmulti}.  There the spectrum of
$\bar q \bar q q q$ mesons was obtained by diagonalizing the one-gluon
exchange interaction.  The light states turned out to be predominantly
$\overline{\qq}\qq$, but other diquark types mixed in as well.  So the
pure diquark model being described here is more radical and more
elementary than the one proposed there.  To learn the real extent of
$\qq$ dominance will require more models and more information from
experiment.  The qualitative ideas explored here are not powerful
enough to fix the overall mass scale of any given sector in QCD. So we
cannot \emph{predict} the existence of (nearly) stable exotic
pentaquarks, or determine whether the $H$-dibaryon is stable.  As was
the case of the large $N_{\rm c}$-dynamics of Jenkins and Manohar,
once a particle like the $\Theta^+$ is found, it sets the scale, and
leads to many interesting predictions.

The predictions that follow from $\qq$-dominance are simple, and
striking.  They capture all the important features of exotic
spectroscopy and provide the conceptual basis of a unified description
of this sector of QCD.
\begin{itemize}
\item  {There should be no (light, prominent) exotic mesons.}

The good diquark, $\qq$, is a flavor $\mathbf{\overline{3}}$, just 
like the antiquark.  Tetraquarks, $\bar 
q\bar q q q$, potentially include exotics in $\mathbf{27}$,  
$\mathbf{10}$, and $\mathbf{\overline{10}}$ representations 
of flavor $SU(3)$.  However $\overline{\qq}\otimes\qq$ contains 
\emph{only non-exotic} representations, $\mathbf{1}$ and 
$\mathbf{8}$, just like $\bar q \otimes q$:
\begin{eqnarray}
q^{{\mathbf 3}}\otimes \bar q^{\mathbf {\overline{3}}}
&=& (\bar q q)^{\mathbf{1}}\oplus  (\bar q q)^{\mathbf{8}}
\nonumber\\
\overline\qq^{{\mathbf 3}}\otimes \qq^{\mathbf {\overline{3}}}
&=& (\overline\qq \qq)^{\mathbf{1}}\oplus (\overline\qq
\qq)^{\mathbf{8}} \nonumber\\
\label{non-exotic}
\end{eqnarray}
Other diquark-antidiquark mesons are heavier, where they would be
buried in the meson-meson continuum.  As described in Section III.C
probably they are not just ``broad'', but in fact absent\cite{jl}.

\item {The only prominent tetraquark mesons should be an 
${SU(3)}$ nonet with ${J^{\Pi}=0^{+}}$}.

This prediction --- a simple corollary of the one just above --- dates
back to the late 1970's\cite{rjmulti}.  Since the diquarks in
eq.~(\ref{non-exotic}) are spinless bosons, the spin$^{\rm parity}$ of
the lightest nonet is $J^{\Pi}=0^{+}$.  Over the years evidence has
accumulated that the nine $0^{+}$-mesons with masses below 1 GeV (the
$f_{0}(600)$, $\kappa(800)$, $f_{0}(980)$, and $a_{0}(980)$) have
important tetraquark components\cite{amsler,close&tornqvist,newlook}. 
Details will follow.

\item{If there are any exotic pentaquark baryons, they lie in
a positive parity ${\overline{10}}$ of ${SU(3)_{\rm 
f}}$.}

This is also a simple consequence of combining good diquarks.  To make
pentaquarks it is necessary to combine two diquarks and an antiquark. 
The result is
\begin{equation} 
\qq^{\mathbf{\overline{3}}} \otimes
\qq^{\mathbf{\overline{3}}}\otimes \bar
q^{\mathbf{\overline{3}}} = (\qq\qq\bar
q)^{\mathbf{1},\frac{1}{2}^{-}}\oplus (\qq\qq\bar
q)^{\mathbf{8},\frac{1}{2}^{-}} \oplus (\qq\qq\bar
q)^{\mathbf{8},(\frac{1}{2}^{+},\frac{3}{2}^{+})} \oplus
(\qq\qq\bar
q)^{\mathbf{\overline{10}},(\frac{1}{2}^{+},\frac{3}{2}^{+})}
\label{pentaquarks}
\end{equation}
The only exotic in $\mathbf{\bar 3}\otimes\mathbf{\bar 3}\otimes
\mathbf{\bar 3}$ is the $\mathbf{\overline{10}}$.  Other exotic flavor
multiplets, like the $\mathbf{27}$ and $\mathbf{35}$, which occur in
the uncorrelated quark picture and/or the chiral soliton models,
should be heavier and most likely lost in the meson-baryon continuum. 
The spin and parity assignments in eq.~(\ref{pentaquarks}) and the
many properties of these pentaquark baryons are discussed below.

\item{ Nuclei will be made of nucleons.}

To a good approximation, nuclei are made of nucleons --- a fact which
QCD should explain.  If diquark correlations dominate, systems of $3A$
quarks should prefer to form individual nucleons, not a single hadron.

The argument is based on statistics: Good diquarks are spinless color
anti-triplet bosons.  Only one, $[u,d]$, is non-strange.  A six-quark
system made of three of these, antisymmetrized in color to make a
color singlet, would have to have fully antisymmetric
space-wavefunction to satisfy Bose statistics.  The simplest would be
a triple-scalar product, $\vec p_{1}\cdot\vec p_{2}\times \vec p_{3}$,
which should be much more energetic than two separate, color-singlet
nucleons in an $s$-wave ({\it eg.\/} the deuteron).  The argument
generalizes to heavy nuclei.  Of course it does not explain nuclear
binding or the rich phenomena of nuclear physics.

\item{The $ H $-dibaryon looks less attractive}

Long ago I argued that perturbative color-spin interactions were
maximally attractive in the $uuddss$ system and might bind a spinless,
doubly strange dihyperon\cite{rljh}.  Searches for the $H$ have so far
come up empty, restricting its binding energy to be less than $\sim$ a
few MeV\cite{limit}.

From a diquark perspective, the $H$ is special: The only way to put
three diquarks in a totally symmetric (low energy) space state
requires one diquark of each flavor, $[u,d][d,s][s,u]$ --- just the
quantum numbers of the $H$.  There is, hidden in this description, a
hint of repulsion in the $H$-system coming from Pauli blocking. 
Although good diquarks are bosons, they are composed of fermions, the
quarks, and each diquark has quark components identical in flavor,
color, and spin with quarks in the other diquarks.  Therefore the
exclusion principle will generate a repulsion between good diquarks,
or equivalently mix other, less favorable diquarks into the ground
state.  Of course the ground state with respect to one-gluon exchange,
which was constructed in Ref.~\cite{rljh} obeys the proper statistics
and therefore does not consist solely of good diquarks.  If multiquark
hadron stability is driven mainly by the good diquark correlation,
there is reason for the $H$ to be less bound.  Our tools are too blunt
to settle the question, which will be decided either by more accurate
theoretical methods or by experiment\cite{hlattice,hexpt}.

\end{itemize}

\subsection{The scalar mesons}
Nearly all known mesons made of $u$, $d$, and $s$ quarks fit neatly
into the multiplets expected in generic constituent quark models.  The
single, striking exception are the scalar ({\it ie\/} $J^{\Pi C}
=0^{++}$) mesons with masses below 1 GeV. The classification of these
mesons has been a bone of contention for more than 30 years.  The
history of the $\sigma$-meson, the broad, isosinglet $\pi\pi$ $s$-wave
resonance near 600 MeV, now known as the $f_{0}(600)$ is a good case
in point: For years listed by the PDG, it was exiled to the gulag of
particle physics in the 1980's, but now has  been rehabilitated and
lives comfortably in the pages of the latest edition of the PDG. The
$f_{0}(980)$ and $a_{0}(980)$ --- twin scalar resonances just at
$\overline KK$ threshold, the first an isoscalar, the second an
isovector --- are well established, but their shapes and
interpretation are complicated by their proximity to and strong
interaction with the $\overline K K$ threshold.  They prefer to couple
to $\overline KK$, a channel with little phase space, instead of
$\pi\pi$ or $\pi\eta$.  Finally, the four light $K\pi$ isospin-$1/2$
resonances near 800 MeV, known as the $\kappa(800)$ or
$K_{0}^{*}(800)$, remain too controversial for inclusion in the latest
PDG Summary Table, although they make an appearance in the ``Particle
Listings''.  Whether or not they are recognized as ``states'' by the
PDG, all these $s$-wave enhancements are obvious in the data, where
they correspond to strong attractive interactions in meson-meson
scattering at low energies.

Altogether these nine states form an anomalously light nonet of
scalar mesons.  $J^{\Pi C}=0^{++}$ $\bar q q$ states are expected
near the other positive parity mesons ($1^{+\pm}\ \&\ 2^{++}$)
between 1200 and 1500 MeV. And over the years a nonet and more
have been found in this region.  Since scalar glueballs are also
expected at these mass scales, this domain has important issues of
its own.  The existence of a scalar nonet above 1 GeV renders the
light scalars supernumerary in a $\bar q q$ classification scheme. 
The $\overline \qq \qq$ interpretation and other possibilities
have been discussed in two recent, thorough reviews\cite{amsler,
close&tornqvist}.

\begin{figure}
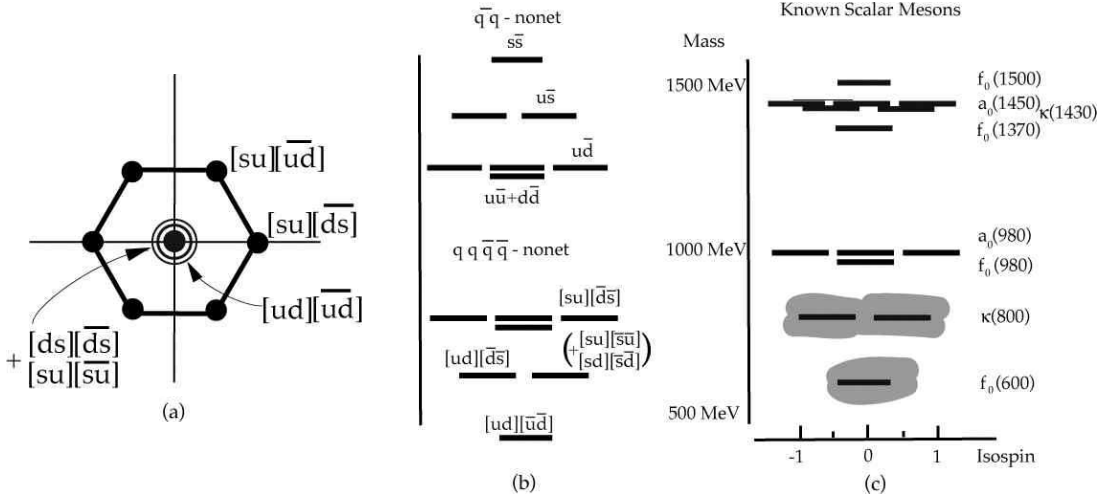

\begin{center}
\BoxedEPSF{scalarnonet-.epsf scaled 360} \qquad
\BoxedEPSF{mesonmasses-.epsf scaled 360} \caption{A $\overline\qq\qq$
nonet: (a) Quark content of a nonet composed of
$\overline\qq\otimes\qq$.  (b) Comparison of the mass spectrum of an
ideally mixed $\bar q q$ nonet (like the vector mesons) and an ideally
mixed $\overline\qq\qq$ nonet.  Note the inversion.  (c) The masses of
the light scalar mesons listed in the Particle Data Tables\cite{pdg}.}
\label{scalarnonet}
\end{center}
\end{figure}

The diquark model of the scalar mesons is quite straightforward.  More
detailed descriptions can be found in Refs.~\cite{maiani,breit} in
addition to the reviews, Refs.~\cite{amsler,close&tornqvist}. 
Briefly: the simplest hadrons made of a scalar diquark and antidiquark
are shown schematically in Fig.~\ref{scalarnonet}.
Fig.~\ref{scalarnonet}(a) shows the quark content of mesons composed
of the diquark $\mathbf{\overline{3}}_{\rm f}$ and the antidiquark
$\mathbf{3}_{\rm f}$ assuming that the strange quark mass effects are
treated to first order.  At this order, the two isoscalars mix
ideally, so one is $\bar u\bar d u d$ and the other is $\bar s s(\bar
u u+\bar d d)$, naturally degenerate with the isovector, $\bar s s\bar
d u$, $\bar s s(\bar uu-\bar d d)$, and $\bar s s\bar u d$. 
Fig.~\ref{scalarnonet}(b) compares the spectrum of $\overline\qq\qq$
mesons with a traditional $\bar q q $ nonet like the vector mesons. 
The $\overline\qq\qq$ spectrum is inverted.  The lightest state is the
non-strange isosinglet ($\bar u\bar d ud$).  The heaviest are the
degenerate isosinglet and isovector which contain ``hidden'' $\bar s
s$ pairs.  The four strange states lie in between.  In contrast, the
$\rho$ and $\omega$ are light and degenerate and the predominantly
$\bar s s$ $\phi$ meson is heavy.  The spectrum of known light-quark
scalar mesons is shown in Fig.~\ref{scalarnonet}(c), taken from the
PDG Tables\cite{pdg} (the smudges denote the very wide $f_{0}(600)$
and $\kappa(800)$).  The similarity between the pattern of the known
light mesons and the $\overline\qq\qq$ states speaks for itself. 
There is much more to be considered: widths, branching ratios, photon
decays, production in $\gamma\gamma$ collisions, {\it etc\/}, all of
which are discussed in Refs.~\cite{amsler} and \cite{close&tornqvist}.

\subsection{Pentaquarks from diquarks I: The general idea}

The diquark picture of pentaquarks follows the same general principles
as the description of tetraquark mesons.  We assume that the scalar
diquark dominates the spectrum.  The rest follows from rather simple
considerations of the symmetry of the $\qq\qq$ wavefunction in color,
flavor, and space (the spin wavefunction is trivial)\cite{jw,sn}.  The
good diquark is a spinless boson so the $\qq\qq$ wavefunction must be
symmetric under interchange of the two diquarks.  The two diquarks
must couple to a color $\mathbf{3}_{\rm c}$:
\begin{equation}
\left[\qq^{\mathbf{\overline{3}}_{\rm c}}\qq^{\mathbf{\overline{3}}_{\rm c}}
\right]^{\mathbf{3}_{\rm c}}
\end{equation}
\begin{figure}
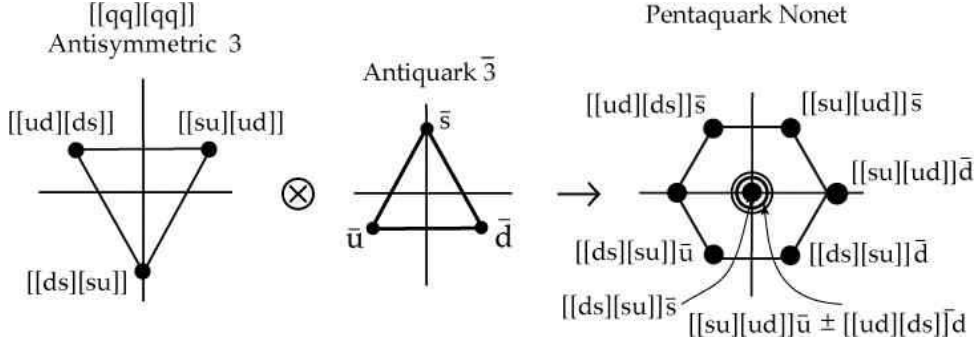

\begin{center}
\BoxedEPSF{oddpenta-.epsf scaled 500} 
\caption{Odd parity pentaquark nonet:  Flavor antisymmetric 
diquarks, $\qq$, in the $\m{\overline 3}_{\rm f}$ representation, are 
combined antisymmetrically, hence the notation $[[qq][qq]]$, and then 
combined with the antiquark, $\m{\overline 3}_{\rm f}$.  The quark 
structure of most of the nonet states is shown assuming ideal mixing.}
\label{figoddpentas}
\end{center}
\end{figure}
so the $\qq\qq$ wavefunction is \emph{antisymmetric} in color.  Two
choices remain: It can be (a) antisymmetric in flavor and symmetric in
space; or (b) symmetric in flavor and antisymmetric in space. 
Symmetric in flavor means $\mathbf{6}$ and antisymmetric means
$\mathbf{\overline{3}}$: $ \left[\mathbf{\overline{3}}
\otimes\mathbf{\overline{3}}\right]_{\cal S} = \mathbf{\overline{6}},
\ \left[\mathbf{\overline{3}}
\otimes\mathbf{\overline{3}}\right]_{\cal A} = \mathbf{3} $.  Symmetry
in space means even parity and a tower of states presumably beginning
with $\ell=0$.  Antisymmetry in space means odd parity and a tower
beginning with $\ell=1$.

So the candidates for light pentaquarks in the diquark scheme fall 
into two categories,

\vspace*{2ex}
\noindent {\bf (a) A negative parity nonet with
$\mathbf{J^{\Pi}=1/2^{-}}$}

\vspace*{.5ex}

The space, color, and flavor structure of the state is 
summarized by,
\begin{equation}
	\left|[\qq\qq]^{\ell=0,\mathbf{3}_{\rm c},\mathbf{3}_{\rm f}}\
	\bar q^{\ j=\frac{1}{2}
	,\mathbf{\overline{3}}_{\rm c},\mathbf{\overline{3}}_{\rm f}}
	\right\rangle^{J^{\Pi}=\frac{1}{2}^{-},
	\mathbf{1}_{\rm c},(\mathbf{1}_{\rm f}\oplus\ \mathbf{8}_{\rm f})}
	\label{oddpentas}
\end{equation}
and the quark content of the nine states is summarized in
Fig.~\ref{figoddpentas}.  In the figure, I assume ideal mixing
({\it ie\/} diagonalizing the number of strange quarks), the
motivation for which is discussed below.

\vspace*{2ex}
\noindent{\bf(b) A positive parity 18-plet (an octet and
antidecuplet) with $\mathbf{J^{\Pi}=1/2^{+}\ \&\  3/2^{+}}$}

\begin{figure}
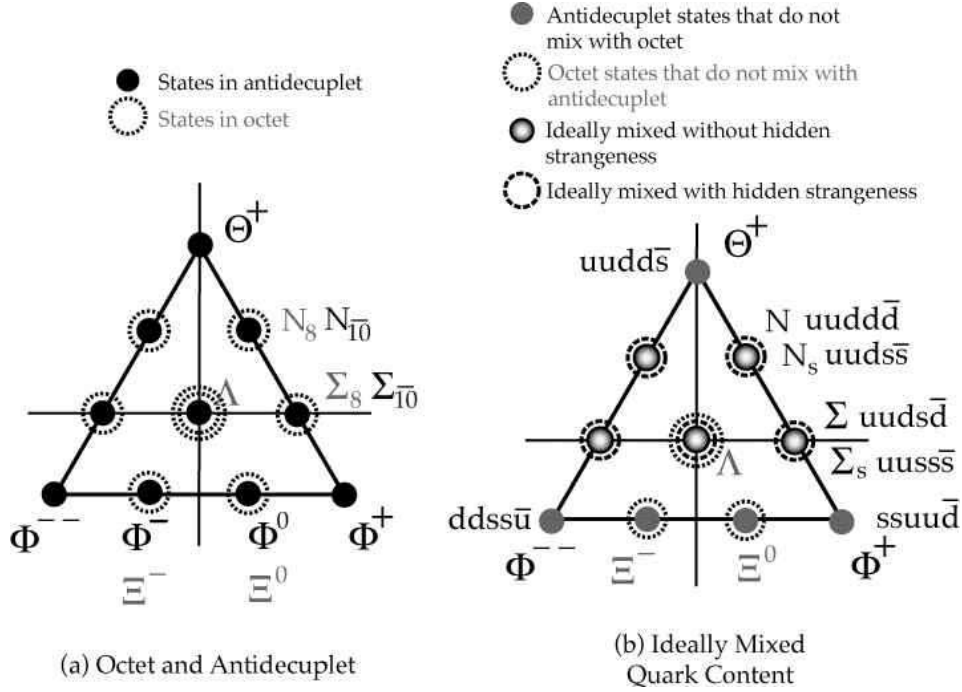

\begin{center}
\BoxedEPSF{evenpenta-.epsf scaled 500} \caption{Even parity
pentaquark 18-plet: diquark pairs in the $\m{\overline 6}_{\rm f}$
combine with an antiquark in the $\m{\overline 3}_{\rm f}$ to make a
$\m{8}_{\rm f}$ and $\m{\overline{10}}_{\rm f}$.  The $SU(3)$ weight
diagram for the $\m{8}_{\rm f}$ and $\m{\overline{10}}_{\rm f}$ is
shown at left, where the unmixed states are named (the decuplet in
black, the octet in grey).  The ideally mixed states, some with their
valence quark content, are shown at right.  The exotics ($\Theta^+$,
$\Phi^{--}$, and $\Phi^{+}$) and certain octet states ($\Lambda$,
$\Xi^{0}$, $\Xi^{-}$) do not mix if isospin is a good symmetry.}
\label{figevenpentas}
\end{center}
\end{figure}

\vspace*{.5ex}
Here the space, color, and flavor structure  is 
summarized by,
\begin{equation}
	\left|[\qq\qq]^{\ell=1,\mathbf{3}_{\rm c},\mathbf{\overline{6}}_{\rm f}}\
	\bar q^{\ j=\frac{1}{2}
	,\mathbf{\overline{3}}_{\rm c},\mathbf{\overline{3}}_{\rm f}}
	\right\rangle^{J^{\Pi}=(\frac{1}{2}^{+}\oplus \frac{3}{2}^{+}),
	\mathbf{1}_{\rm c},(\mathbf{8}_{\rm f}\oplus\ \mathbf{
	\overline{10}}_{\rm f})}
	\label{oddpentas1}
\end{equation}
and the quark content of the eighteen states is summarized in
Fig.~\ref{figevenpentas}, where, as in the previous case, ideal mixing
can be assumed.  The figure deserves careful study: the $SU(3)_{\rm
f}$ weight diagrams of the unmixed octet and antidecuplet are shown on
the left.  The results after ideal mixing are shown on the right.  The
exotics in the antidecuplet do not mix with the octet.  Isospin
symmetry precludes mixing between the $\Lambda$ and the $\Sigma^{0}$s
or between the $\Xi^{0,-}$ and the $\Phi^{0,-}$.  The other states,
the $N$s and the $\Sigma$s, mix to diagonalize the number of $\bar s
s$ pairs.  One set has hidden strangeness, the other does not.

It is straightforward to construct the explicit wavefunctions for all
these states using the Clebsch-Gordan coefficients for angular
momentum and for the symmetric and antisymmetric combinations of color
and flavor triplets.  There is a small subtlety concerning the phases
of the diquark antitriplet states, which is correctly implemented in
eq.~(\ref{correspondences})\cite{rlj-unpubl}.  The wavefunctions can
be found written out in more detail, for example, in
Ref.~\cite{close2}.

Which multiplet, the odd parity nonet or the even parity 18-plet, is
lighter depends on the quark model dynamics.  This is exactly the same
question we encountered in the large $N_{\rm c}$ classification of Jenkins
and Manohar.  Color-spin interactions modeled after one gluon exchange
(see eq.~(\ref{colorspin})) favor the $s$-wave, {\it ie\/} the odd
parity nonet, but there may be Pauli blocking in this state as in the
$H$.  This effect would elevate the mass of the negative parity 
nonet. 
Flavor-spin interactions, modeled after pseudoscalar meson
exchange\cite{riskastancu}, apparently favor the $p$-wave (in contrast
to the $\bar q q$, $qqq$, and $\bar q \bar q q q$ sectors where the
ground state is always the $s$-wave), making the even-parity 18-plet
the lightest.  Whichever way, the diquark picture leads to clear
predictions for the light pentaquarks:
\begin{itemize} 
	\item The only potentially light, prominent exotic multiplet is
	the antidecuplet, which contains candidates for the $\Theta^{+}$,
	the $\Phi^{--}$, and an as yet unreported $\Phi^{+}$.
	
	\item The exotics are accompanied by an non-exotic octet, which
	mixes with the antidecuplet to give several non-exotic (or
	``cryptoexotic'') analogue states, for example a $[u,d][u,d]\bar u$
	and $[u,d][u,d]\bar d$ pair, which should be lighter than the
	$\Theta^+$.
	
	\item  There are no other light, prominent exotics, like the 
$\mathbf{27}$
	that figures prominently in the chiral soliton model.
	
	\item The $\Theta^+$ should have positive parity.
	
	\item The exotics should come in spin-orbit pairs with 
	$J^{\Pi}=\frac{1}{2}$ and $\frac{3}{2}$.
\end{itemize}
More predictions include $SU(3)_{\rm f}$ mass splittings and the 
existence
of charm and bottom analogue states discussed below.
	
I will have little further to say about the negative parity nonet. 
These states couple strongly to the meson-nucleon $s$-wave.  The
non-strange members of the multiplet contain an $\bar s s$ pair and
should therefore couple to $N\eta$ and $\Lambda K$, not to $N\pi$. 
Unless these states were below fall apart decay threshold they would
be lost in the meson-nucleon continuum.  The absence of candidates in
the PDG tables should not be surprising.

A word about complications that I have ignored in this presentation:
First are the states constructed from the other diquarks: Residual
interactions will certainly mix them into the $\qq\qq\bar q$ states,
but at zeroth order the good$\times$bad and bad$\times$bad states are
$\sim$200 and $\sim$400 MeV heavier than the good$\times$good states. 
If the lightest states in each family are the $s$-waves --- as QCD
based interactions prefer --- then these states are all well above
threshold to fall apart into meson and baryon, and disappear into the
continuum.  Among them are many exotics, but only one candidate for a
negative parity antidecuplet, consistent with our earlier discussion
(see eq.~(\ref{quarktenbar})).  Good$\times$bad states lie in the
$\mathbf{\overline{3}}\otimes \mathbf{6} \otimes
\mathbf{\overline{3}}$ of $SU(3)_{\rm f}$ which includes the exotic
$\mathbf{27}$.  Bad$\times$bad states lie in the $\mathbf{6} \otimes
\mathbf{6} \otimes \mathbf{\overline{3}}$ and include a negative
parity antidecuplet (as well as the $\mathbf{35}$).  So the first
candidate for a \emph{negative parity} $\Theta^+$ lies in the
``bad-bad'' sector and furthermore is created by the same operator
that creates $KN$ in an $s$-wave.  So the diquark picture is quite
firm that a \emph{negative parity} $\Theta^+$ is much heavier and
strongly coupled to the $KN$ $s$-wave continuum.

Second is mixing between $qqqq\bar q$ states and ordinary $qqq$ 
baryons.  Mixing is possible when the $qqqq\bar q$ states are not 
exotic, especially if there are $qqq$ states with the same quantum 
numbers nearby.  Mixing will alter both the spectrum and the decay 
widths that would otherwise be determined by $SU(3)$ flavor symmetry.

\subsection{Pentaquarks from diquarks II:  \\ A more detailed look at 
the
positive parity octet and antidecuplet}
	 
If the $\Theta^+$ and its brethren are confirmed, and if they have
positive parity, then the diquark based pentaquark picture seems like
a strong candidate for a quark description of the structure.  This
section and the next are devoted to describing the predictions of the
diquark picture in some detail, as presented in Refs.~\cite{jw,
jw2,jw3}

The states of the positive parity 18-plet are labeled in
Fig.~\ref{figevenpentas}, with names assigned according to the new PDG
conventions: $Y=2$, $I=0 \ \Rightarrow \Theta$, $Y=-1$, $I=3/2 \
\Rightarrow \Phi$; and in the case of residual ambiguity, by appending
a subscript ``$s$'' to states with hidden $\bar s s$ pairs.

\subsubsection{Flavor $SU(3)$ violation and mass relations}

The standard approach to incorporating $SU(3)$ violation in quark
spectroscopy is to include the effects of the strange quark mass to
lowest order in perturbation theory, which has been perfectly adequate
for all $qqq$ baryons and $\bar q q$ mesons in the past.  The
perturbing hamiltonian, ${\cal H}'$, is therefore proportional to the
$SU(3)_{\rm f}$ hypercharge.  There is no {\it a priori\/} reason to
expect it to fail for pentaquarks.  At this point there is no reason
to assume ideal mixing, so I refer to the $Y=1$ mass eigenstates as
$N$ and $N'$ and the $Y=0$, $I=1$ mass eigenstates as $\Sigma$ and
$\Sigma'$.  This gives eight masses to be fit (in order of decreasing
strangeness): $\Theta^+$, $N$, $N'$, $\Lambda$, $\Sigma$, $\Sigma'$,
$\Xi$, and $\Phi$ (assuming no isospin violation).  The parameters of
the fit include: the unperturbed octet and antidecuplet masses,
$M_{\mathbf 8}$, and $M_{\mathbf{\overline{10}}}$, and the reduced
matrix elements of the perturbing Hamiltonian:
\begin{equation}
	\langle \m{\overline{10}}\parallel{\cal H}'\parallel
	\m{\overline{10}}\rangle,\ \ \langle
	\m{\overline{10}}\parallel{\cal H}'\parallel \m{8}\rangle,\ \
	\langle \m{8}\parallel{\cal H}'\parallel \m{8}\rangle_{D},\ \
	\langle \m{8}\parallel{\cal H}'\parallel \m{8}\rangle_{F}\ \
\end{equation}
where the subscripts $F$ and $D$ refer to the usual duplication of
$\mathbf{8}$ in $\mathbf{8}\otimes\mathbf{8}$.  Six parameters and
eight masses leave two mass relations.  One is hopelessly non-linear,
the other,
\begin{equation}
	2(N+N'+\Xi)=\Sigma+\Sigma'+3\Lambda +\Theta
\end{equation}
was, to my knowledge, first written down by Diakonov and
Petrov\cite{dpmass}.  It can be violated if the pentaquark states mix
with other, {\it eg\/} $qqq$, multiplets nearby.  For example the
pentaquark ``Roper'' can mix with a radially excited nucleon. 

It is possible to make more progress by imposing some of the structure
suggested by the one-gluon exchange motivated interaction,
eq.~(\ref{colorspin}).  The Hamiltonian of eq.~(\ref{colorspin}) is
color and spin dependent, and depends on the flavor of the quarks
explicitly through $M_{ij}$ and implicitly via fermi statistics which
correlates flavor with color and spin in states of definite space
symmetry.  However, ${\cal H}_{\rm color\ spin}$ does not distinguish
between the $\mathbf{8}_{\rm f}$ and $\mathbf{\overline{10}}_{\rm f}$
in the ``final'' flavor coupling of
$\qq\qq^{\mathbf{\overline{6}}_{\rm f}}$ to $\bar q^{
\mathbf{\overline{3}}_{\rm f}}$.  Furthermore the $SU(3)$ breaking in
the kinetic and confining pieces of the Hamiltonian simply counts the
number of $s$-quarks.\footnote{ These observations do not apply to
other pictures of the residual quark-quark interactions, like the
flavor-spin picture of Refs.~\cite{riskastancu}, where different
patterns of $SU(3)_{\rm f}$ symmetry violation arise.} As a result, a)
the octet and antidecuplet are degenerate in the absence of
$SU(3)_{\rm f}$ symmetry violation, and b) the symmetry breaking
Hamiltonian, ${\cal H}'$ acts in the $\qq\qq$ and $\bar q$ sectors
independently.  This leaves only three parameters,
\begin{equation}
	M_{\m{8}} = M_{\m{\overline{10}}}, \  
	\langle\{\qq\qq\}\m{\overline{6}_{\rm f}}\parallel{\cal H}'\parallel
	  \{\qq\qq\}\m{\overline{6}_{\rm f}}\rangle,  \
	  \langle\{\bar q\}\m{\overline{3}_{\rm f}}\parallel{\cal 
H}'\parallel
	  \{\bar q\}\m{\overline{3}_{\rm f}}\rangle 
\end{equation}
and considerably more predictive power.  In Ref.~\cite{jw} Wilczek and
I chose three different parameters which are linear combinations of
these,
\begin{eqnarray}
	 M_{0} &=&M_{\m{8}}+\frac{4}{3}
	 \langle\{\qq\qq\}\m{\overline{6}_{\rm f}}\parallel{\cal H}'\parallel
	  \{\qq\qq\}\m{\overline{6}_{\rm f}}\rangle-\frac{1}{3}
	  \langle\{\bar q\}\m{\overline{3}_{\rm f}}\parallel{\cal H}'\parallel
	  \{\bar q\}\m{\overline{3}_{\rm f}}\rangle  \nonumber\\
	 \mu &=&\langle\{\bar q\}\m{\overline{3}_{\rm f}}\parallel{\cal H}'\parallel
	  \{\bar q\}\m{\overline{3}_{\rm f}}\rangle\nonumber\\
	  \alpha &=&-\langle\{\bar q\}\m{\overline{3}_{\rm
	  f}}\parallel{\cal H}'\parallel \{\bar
	  q\}\m{\overline{3}_{\rm f}}\rangle-
	  \langle\{\qq\qq\}\m{\overline{6}_{\rm f}}\parallel{\cal
	  H}'\parallel \{\qq\qq\}\m{\overline{6}_{\rm f}}\rangle
\end{eqnarray}
where $\mu$ is the matrix element of $m_{s}\bar s s$ and $\alpha$ is
$M[u,s]-M[u,d]$.\footnote{In Ref.~\cite{jw} $\mu$ was denoted as
$m_{s}$ engendering some confusion.} In terms of these,
\begin{eqnarray}
	M(N) &=& M_{0}\nonumber\\
	M(\Theta)&=& M_{0} +\mu \nonumber\\
	M(\Lambda)=M(\Sigma) &=& M_{0}+\mu+\alpha \nonumber\\
	M(N_{s}) &=& M_{0} + 2\mu +\alpha\nonumber\\
	M(\Phi)=M(\Xi)&=& M_{0} +2\mu  +2\alpha\nonumber\\
	M(\Sigma_{s}) &=& M_{0} +3\mu+2\alpha
\end{eqnarray}

To pin down the spectrum it is necessary to identify some of these
$\qq\qq\bar q$ states with physical hadrons or import values of the
parameters from elsewhere in the hadron spectrum.  Before the report
of the $\Phi^{--}$, Wilczek and I used the $\Theta^+$ with mass 1540
MeV and the Roper --- identified with the $N$ --- at 1440 MeV, to
extract $\mu\approx 100$ MeV. We took the parameter $\alpha \approx
60$ MeV from the quark model analysis of the $\Lambda$-$\Sigma$ mass
difference.  The resulting predictions for the masses of the 18-plet
--- together with possible candidates --- are listed in the fourth
column (labelled ``Mass I'') of Table~\ref{assignments}.

\begin{table}[htdp]
\caption{Masses and possible assignments of 18-plet pentaquark states 
in the diquark picture.}
\begin{center}
\begin{tabular}{|l|l|c|c|c|c|c|}
\hline
Name & Quark content &Mass &Mass I (MeV)\cite{jw} &Mass II
(MeV)\cite{jw3}&$\matrix{\mbox{Possible}\cr \mbox{Assignment}}$&
Comments\\
\hline
$N$ &$[ud][ud]\bar u$, \ldots &\ $M_{0}$ & 1440 & 1450 & 
$P_{11}(1440)$ Roper &
$\matrix{\mbox{Width is problematic}\cr \mbox{Input for II}}$\\
\hline
$\Theta^+$ &$[ud][ud]\bar s$&\ $M_{0}+\mu $ & 1540 & 1550 & 
$P_{01}(1540)$ 
$\Theta^+$ & Input for II\\
\hline
$\Lambda$, $\Sigma$ &$[ud][su]\bar d$, \ldots&\ $M_{0}
+\mu+\alpha$ & 1600 & 1650 & $P_{13}(1660)\ \Sigma$  & Input for II
 \\
\hline
$N_{s}$ &$[ud][su]\bar s$, \ldots&\ $M_{0}+2\mu+\alpha$
& 1700 & 1750 & $P_{11}(1710)$ &  \\
\hline
$\Phi$, $\Xi$ &$[su][su]\bar d$, \ldots&\ $M_{0}+2\mu +2\alpha $ &
1760 & 1850 & $\matrix{ \Phi^{--}(1860) \cr  \Xi^{-}(1860)}$ &\\
\hline 
$\Sigma_{s}$ &$[su][ds]\bar s$, \ldots&\ $M_{0} +3\mu +2\alpha $ &
1860 & 1950 &?  &\\
\hline 
\end{tabular}
\end{center}
\label{assignments}
\end{table}%
At the time the prediction of a relatively light $\Phi$ was rather
daring.  Now that the $\Phi^{--}$ has reported at 1860 MeV, it is
appropriate to re-examine these predictions and assignments.  In
retrospect taking $\alpha$ from the $\Lambda$-$\Sigma$ system may not
have been particularly appropriate.  In the notation of Section IV,
$\alpha = M[u,s]-M[u,d]$, which cannot be extracted reliably from any
measured baryon mass differences.  The value extracted in
Ref.~\cite{jw} assumed the full color-spin Hamiltonian of
eq.~(\ref{colorspin}).  Instead in Ref.~\cite{jw3} Wilczek and I
propose to identify $\Sigma(1660)$, a $1/2^{+}$ resonance given three
stars by the PDG\cite{pdg}, with the $\Sigma$ states in the 18-plet. 
This choice is motivated by a global fit to baryon
resonances\cite{sw}.  We are also mindful that the $N(1710)$ is a
candidate for the $N_{s}$.  Given the widths of these states (the
Roper alone has a width of 350 MeV), it makes no sense to quote masses
to an accuracy greater than, say, 50 MeV. The resulting alternative
mass spectrum is shown in the fifth column of Table~\ref{assignments}.

There is an important, qualitative difference between the diquark
picture with its 18-plet and other models of the $\Theta^+$ with an
antidecuplet alone, which will help sort out the correct physical
picture of the exotics.  $SU(3)$-flavor splittings within the
antidecuplet obey an equal spacing rule, just like the better known
decuplet ($\Delta$, $\Sigma^{*}$, $\Xi^{*}$, $\Omega^{-}$) where it is
very successful.  In an antidecuplet-only picture the $\Theta^+$,
$N_{\m{\overline{10}}}$, $\Sigma_{\m{\overline{10}}}$, and $\Phi$ must
be spaced at equal mass intervals.  In their original paper\cite{dpp}
Diakonov, Petrov, and Polyakov identify the $N_{\m{\overline{10}}}$
with the $N(1710)$, which puts the $\Phi$ at 2070 MeV, much higher
than the quark content ($uudd\bar s \ \to ddss\bar u$) would
suggest\cite{jw}.  If the $\Phi$ lies at 1860, the
$N_{\m{\overline{10}}}$ and $\Sigma_{\m{\overline{10}}}$ must lie near
1650 MeV and 1750 MeV respectively.  The $N(1710)$ is an imperfect
candidate for the first, but there is no candidate for the second
excepting a dubious (one-star) $P_{11}$ at 1770.  In their revised
discussion of the spectrum, Diakonov and Petrov, fit the $\Phi(1860)$
and predict $1/2^{+}$ $N$ and $\Sigma$ resonances in the intervals
1650-1690 and 1760-1810 respectively\cite{dpmass}.  Weigel has
considered mixing of the $\m{\overline{10}}$ with (non-degenerate)
radially excited octets and other exotics like $\m{27}$ and
$\m{35}$\cite{weigel,weigelmix}, and Diakonov and Polyakov mix the
$\m{\overline{10}}$ with the ground state $\m{8}$\cite{dpmass}. 
Others have proposed non-linear $SU(3)_{\rm f}$ violation, but this
seems {\it ad hoc\/} given the lack of similar effects elsewhere in
the spectrum\cite{prasz}.

\begin{figure}
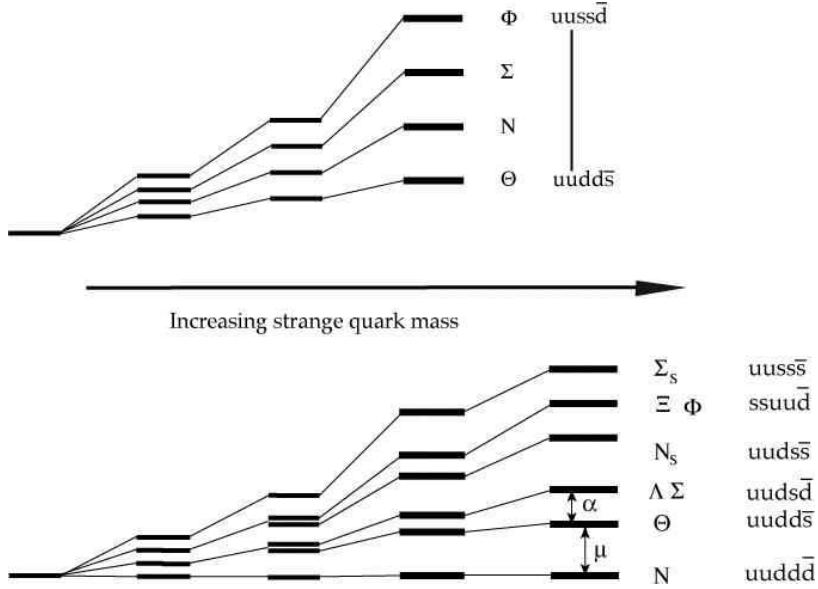

\begin{center}
\BoxedEPSF{splittings-.epsf scaled 450} \caption{The $SU(3)_f$ 
splittings among light $\qqq$ states as a function of $m_s$.  The antidecuplet
-only case on top.  The 18-plet case below.}
\label{splittings}
\end{center}
\end{figure}

In contrast with the $\m{\overline{10}}$-only, the 18-plet picture
suggested by diquark arguments allows the $\Theta^+$ and $\Phi$ to be
interior to the multiplet, with the $\Sigma_{s}(uuss\bar s)$ and
$N(uudd\bar u)$ at the top and bottom respectively. The spectrum of the 
diquark 18-plet picture is contrasted with the antidecuplet-only spectrum in Fig.
~\ref{splittings}

 As explained
above there are possible candidates for all the 18-plet
states.\footnote{Although the \emph{width} of the Roper presents a
problem\cite{cohenroper}, suggesting that non-exotic $qqqq\bar q$
states may mix significantly with $qqq$ states.} Time will tell which
of these qualitatively different spectra are closest to Nature ---
provided, of course, the exotics survive the next round of
experiments.

\subsubsection{Isospin and $SU(3)$ selection rules}

The existence of \emph{two} nearly degenerate isomultiplets of
strangeness minus two pentaquarks (the $\Xi$ and the $\Phi$) is one of
the most robust and striking predictions of diquark picture.  The
degeneracy is exact in the ideal mixing limit described above. 
Isospin conservation prevents the $\Xi^{-,0}$ and $\Phi^{-,0}$ from
mixing, unless they are degenerate within a few MeV (the scale of
typical hadronic isospin violation).  There are interesting
predictions for the decays of these states based on $SU(3)$-flavor
selection rules\cite{jw3,mehen}.  For example, $SU(3)_{\rm f}$ forbids
the decay of the $\Phi$ into a pseudoscalar meson and a member of the
baryon decuplet ($\m{\overline{10}}\to\!\!\!\!\!\!/ \
\m{10}\otimes\m{8}$).  The NA49 Group, which has observed the
$\Phi^{--}$, can only detect charged particles.  They have no evidence
for the exotic $\Phi^{+}$, even though they would be sensitive to
$\Phi^{+}\to\Xi^{*0}(1530)\pi^{+}$.  This is consistent with the
selection rule because the $\Phi^{+}$ must be in the
$\m{\overline{10}}$ and the $\Xi^{*0}(1530)$ lies in the $\m{10}$.  So
$SU(3)_{\rm f}$ seems to be working here.  On the other hand, they do
have (weak) evidence for a $S=-2$, $Q=-1$ state at 1860 MeV decaying
into $\Xi^{*0}(1530)\pi^{-}$.  If this stands up, it identifies this
state as the $\Xi^{-}$, not the $\Phi^{-}$ (since $\m{8}\to
\m{10}\otimes\m{8}$ is allowed) and supports the existence of
degenerate $\Xi$ and $\Phi$ pentaquarks.  For a more detailed tour of
the decay selection rules, see Refs.~\cite{jw3} and \cite{mehen}.
 
\subsection{Pentaquark from diquarks III:   Charm and bottom 
analogues}

Charm and bottom analogues of the $\Theta^+$ can be obtained by 
substituting the heavy $\bar c$ or $\bar b$ quark for the $\bar s$ in 
the $\Theta^+$,
\begin{equation}
	\Theta^{0}_{\rm c}=|[u,d][u,d]\bar c\rangle\qquad\ \ 
	\Theta^{+}_{b}=|[u,d][u,d]\bar b\rangle
	\label{heavyanalogs}
\end{equation}
Charm pentaquark exotics\footnote{The states predicted by Lipkin and
Gignoux {\it et al\/} were not analogues of the $\Theta^{+}$.  They
have the quantum numbers of the states discussed by Stewart, Wessling,
and Wise\cite{sww}.}were predicted many years ago in quark models with
flavor dependent interactions, but were not taken very seriously at
the time\cite{jmr}.  The existence of the
$\Theta^{+}_{s}\equiv\Theta^{+}(1540)$ fixes the mass scale for
exotics and leads to rather robust predictions of the masses of the
$\Theta^{0}_{\rm c}$ and $\Theta^{+}_{b}$.  The simplest, though not
necessarily the least accurate approach, is to find an analogy among
$qqQ$ baryons and apply it to the $\qq\qq\overline Q$ system.  The
obvious choice is the $\Lambda_{Q}$, which has the quark content $\qq
Q$.  The heavy anti-quark in the $\Theta_{Q}$ sits in the background
of an isosinglet, color $\m{3}$, spin singlet pair of diquarks.  The
heavy quark in the $\qq Q$ sits in a background \emph{identical} in
isospin, color, and spin.  The only difference is that the spin of the
$\overline Q$ in the $\Theta_{Q}$ can interact with the orbital
angular momentum ($\ell=1$) in the $\Theta_{Q}$ and this interaction
is not present in the $\Lambda_{Q}$.  Were it not for this, one would
expect the relations
\begin{eqnarray}
M(\qq\qq \bar c)-M(\qq\qq\bar s) = M(\qq c)-M(\qq s) \quad&\mbox{that
is}&\quad M(\Theta^{0}_{\rm c})- M(\Theta^{+}_{s})= M(\Lambda_{\rm c})-
M(\Lambda)\nonumber\\
M(\qq\qq \bar b)-M(\qq\qq\bar c) = M(\qq b)-M(\qq c) \quad&\mbox{that
is}&\quad M(\Theta^{+}_{b})- M(\Theta^{0}_{\rm c})= M(\Lambda_{b})-
M(\Lambda_{\rm c})\nonumber\\
\label{scalingup}
\end{eqnarray}
to be nearly exact.  QCD spin-orbit interactions are not strong, so
these rules should not be badly violated.  Because these interactions
vanish as $m_{Q}\to\infty$, the second relation should be quite
accurate.  The differences among the predictions of various QCD based
quark models reflect the different ways that the residual interactions
are treated.  Taking eqs.~(\ref{scalingup}) as is, Wilczek and I
estimated,
\begin{equation}
	M(\Theta^{0}_{\rm c}) = 2710\ \mbox{MeV} \quad\mbox{and}
	\quad M(\Theta^{+}_{b}) = 6050\ \mbox{MeV}
	\label{heavymasses}
\end{equation}

If these estimates are correct, the $\Theta^{0}_{\rm c}$ and
$\Theta^{+}_{b}$ will be stable against strong decay.  The lightest
strong decay channel for the $\Theta^{0}_{\rm c}$ is $ND$ with a threshold
at 2805 MeV, and for the $\Theta^{+}_{b}$, it is $NB$ with a threshold
at 6220 MeV. They would have to decay weakly with lifetimes of order
$10^{-12}$ sec.

How did this happen?  The $\Theta^{+}_{s}$ is light, but it is not
stable.  The reason lies not in the linear scaling of the masses of
the heavy pentaquarks with the heavy quark mass, but rather in the
\emph{non-linear} scaling of the pseudoscalar meson masses, which
determine the strong decay thresholds.  Consider the four analogue
states, $[u,d][u,d]\overline Q$, with $Q=u,s,c,b$, and identify the
$\Theta^{0}_{u}$ with the Roper as I advocated earlier.  Then
\begin{eqnarray}
	\Theta^{0}_{u}\to N\pi\quad&\mbox{has decay ${\cal Q}$-value}&\quad
	{\cal Q}\approx 350\ \mbox{MeV}\nonumber\\
	\Theta^{+}_{s}\to NK\quad&\mbox{has decay ${\cal Q}$-value}&\quad 
{\cal
	Q}\approx 100\ \mbox{MeV}\nonumber\\
	\Theta^{0}_{\rm c}\to ND\quad&\mbox{has decay ${\cal Q}$-value}&\quad 
{\cal
	Q}\approx -100\ \mbox{MeV}\nonumber\\
	\Theta^{0}_{b}\to NB\quad&\mbox{has decay ${\cal Q}$-value}&\quad
	{\cal Q}\approx -150\ \mbox{MeV}
	\label{qvalues}
\end{eqnarray}
The $\Theta^{0}_{u}$, {\it ie\/} the Roper, is unstable because the
pion is anomalously light, a consequence of approximate chiral
symmetry.  The effect is still significant enough for the kaon to make
the $\Theta^{+}_{s}$ unstable.  The $D$ and $B$-meson masses are not
significantly lowered by chiral symmetry, the thresholds are
proportionately higher, and the $\Theta^{0}_{\rm c}$ and $\Theta^{+}_{b}$
are stable.  The details are model dependent.  Other model estimates
are generally higher than the simple scaling law described
here\cite{Maltmanheavy}, some predict stable $c$ and $b$-exotics,
others predict light and narrow, but not stable states.

For the record, here is a list of some of the more obvious (Cabibbo
allowed) weak decay modes,
\begin{eqnarray}
	\Theta^{0}_{\rm c}&\to& p K^{+} \pi^{-}, \ pK_{S}\pi^{-},\ 
	\Theta^{+}_{s}\pi^{-},\ldots\nonumber\\
	\Theta^{+}_{b}&\to & p \overline{D}^{0},\ p\pi^{+}D^{-},\ 
	\Theta^{0}_{\rm c}\pi^{+},\ p J/\psi K_{S}, \ p J/\psi K^{+}\pi^{-},\
	\Theta^{+}_{s}J/\psi\ (!)\ldots
\end{eqnarray}
and many more.

When the light antiquark is replaced by a charm or bottom, the
previously \emph{non-exotic}, negative parity baryons of
eq.~(\ref{oddpentas}) and Fig.~\ref{figoddpentas} become
exotic\cite{sww,jmr}.  They form a $\m{\overline{3}}$ of $SU(3)_{\rm
f}$: with valence quark content $\{[u,d][d,s]\overline Q$,
$[d,s][s,u]\overline Q$, $[s,u][u,d]\overline Q\}$.  If the $s$-wave
pentaquarks are lighter than the $p$-wave, then these charm and bottom
exotics would be even more tightly bound than the
$\Theta^+$-analogues.  With such strong attraction that the ground
states are stable, it is not surprising that models predict bound or
at least very narrow excited states as well \cite{Maltmanheavy}.

The exotic charm baryon reported by H1 is not bound.  With a mass of
3099 MeV, it is much too heavy to be the $\Theta^{0}_{\rm c}$ as I
have described it.  The width is reported to be less than 12 MeV. It
has been observed through its strong decay into $D^{*-}p$, into which
it has a ${\cal Q}$ value of 150 MeV. If it were the $\Theta^{0}_{\rm
c}$, and if it has $J^{\Pi}=1/2^{+}$, it would have a significant
decay into $D^{-}p$ (which would not have been seen at H1), with a
partial width that can be related to the width of the $\Theta^{+}_{s}$
by scaling $p$-wave phase space.  The result is
$\Gamma(3099)/\Gamma(\Theta^{+}_{s}) \approx 15$, barely consistent
with the H1 limit if the width of the $\Theta^{+}_{s}$ is 1 MeV. An
interesting possibility --- if the 3099 state should be confirmed ---
is that it is an $L=2$ Regge excitation of the $\Theta^{0}_{\rm c}$
with $J^{\Pi}=3/2-$.  This object can decay into $D^{*-}p$ in the
$s$-wave, but $D^{-}p$ in the $d$-wave, accounting perhaps for its
surprisingly narrow width.  Should this assignment prove correct,
there must be many other excited charm exotic baryons awaiting
discovery.

Clearly, if the initial reports are confirmed, there is a fascinating
spectroscopy of heavy exotic baryons awaiting us.  But it is a big
``if''!

\subsection{A paradigm for spectroscopy}

If diquarks are as important in exotic spectroscopy as I have
suggested, their role in ordinary meson and baryon spectroscopy should
be re-examined.  For decades it has been traditional to classify
hadron resonances according to the rules of the non-relativistic quark
model: essentially $SU(N_{\rm f})\times O(3)$.  The latest checklist
can be found in the 2004 Particle Data Tables\cite{pdg}.  To the
extent that diquarks dominate the structure of hadrons, baryons are
more like mesons, $q^{\m{3}_{\rm c}}
{-\!-}\{qq\}^{\m{\overline{3}}_{\rm c}}$ in analogy to $q^{\m{3}_{\rm
c}}{-\!-}\bar {q}^{\m{\overline{3}}_{\rm c}}$, than the three-body
bound states implicit in the quark model classification.  Meson
quantum numbers \emph{and masses} can be understood using a mix of
ideas from QCD and Regge theory: quarks and antiquarks on the ends of
flux-tubes (or strings) with a spectrum determined by the
Chew-Frautschi formula, $M^{2}=\sigma L + \alpha$, where $\sigma$ is
the universal slope of Regge-trajectories (the ``string tension''). 
In a diquark paradigm baryons should lie on trajectories with the
\emph{same slope}, and furthermore group into families depending on
the nature of the diquark (``good'' versus ``bad'') and the coupling
of its spin to the orbital excitation.  This classification program
has been carried out by Selem and Wilczek\cite{sw}, and leads to a
compact and predictive unified picture of mesons, baryons, and
tetraquarks.  One of their observations is that the diquark
correlation appears to become stronger in Regge-extended hadrons.  In
particular, the diquark-quark mass difference, which we called
$M[u,d]-M[u]$, becomes smaller --- evidence that the diquark
correlation is medium dependent.  This is a rich subject, but well
beyond the scope of this review.

\section{Conclusions}
\setcounter{equation}{0}  
There are two distinct, but related issues at the core of this
discussion: first, a question: are there light, prominent exotic
baryons, and if so, what is the best dynamical framework in which to
study them?  and second, a proposal: diquark correlations are
important in QCD spectroscopy, especially in multiquark systems, where
they account naturally for the principal features.  

I believe the case for diquarks is already quite compelling.  There
are many projects ahead: re-evaluating the $qqq$ spectrum\cite{sw};
systematically exploring the role of diquarks in deep inelastic
distribution and fragmentation functions, and in scaling violation;
seeing if diquarks can help in other areas of hadron phenomenology
like form-factors, low $p_{T}$ particle production, and polarization
phenomena; developing a more sophisticated treatment of quark
correlations, recognizing that diquarks are far from pointlike inside
hadrons; establishing diquark parameters and looking for diquark
structure in hadrons using lattice QCD; and --- the holy grail of this
subject --- seeking a more fundamental and quantitative
phenomenological paradigm for light quark dynamics at the confinement
scale.  Diquark advocates have considered many of these issues in the
past\cite{diquarks}.  No doubt many other important contributions,
like the diquark analysis of the $\Delta I=1/2$-rule\cite{neubert},
have already been accomplished.  We can hope eventually to have as
sophisticated an understanding of diquark correlations as we have of
$\bar q q$ correlations, as expressed in chiral dynamics.

The situation with the $\Theta^+$ is less clear.  Of course it will
eventually be clarified by experiment --- a virtue of working on QCD
as opposed to string theory!  However, theorists' attempts to
understand the $\Theta^+$ have raised more questions than they have
answered.  To wit:  
\begin{itemize}
	\item A negative parity ($KN$ $s$-wave) $\Theta^+$ is intolerable to
	theorists, but that is what lattice studies find, if they find
	anything at all.  
	
	\item No one has come up with a simple, qualitative explanation for 
	the exceptionally narrow width of the $\Theta^+$.
	
	\item The original prediction of a narrow, light $\Theta^+$ in the
	chiral soliton model does not appear to be robust.
	
	\item Quark models can accomodate the $\Theta^+$, but only by
	reversing the naive, and heretofore universal, parity of the
	$q^{n_{q}}\bar q^{n_{\bar q}}$ ground state.  It is necessary to
	excite the quarks in order to capture the correlation energy of
	the good diquarks.  This does not sound like a way to make an
	exceptionally light and stable pentaquark.
	
	\item When models are adjusted to accomodate the $\Theta^+$, they
	predict the existence of other states that should have been
	observed by now: The diquark picture wants both a
	$\Theta^{\frac{1}{2}^{+}}$ and a $\Theta ^{\frac{3}{2}^{+}}$; the
	CSM and large $N_{\rm c}$ want a relatively light $\m{27}$, which
	includes an $I=1$ triplet: $\Theta^{*0},
	\Theta^{*+},\Theta^{*++}$.
\end{itemize}
None of these problems seems insuperable.  Indeed, there are papers
appearing every day that propose an interesting solution to one or
another.  Taken together, however, they are an impressive set.  They
leave us in limbo: Either the $\Theta^+$ will go away, or it will
force us to rewrite several chapters of the book on QCD.

\section{Acknowledgements}

Many of these ideas were developed in collaboration with Frank
Wilczek.  I have also benefited from conversations, correspondence and
collaborations with Tome Anticic, Pat Burchat, Carl Carlson, Frank
Close, Tom Cohen, Jozef Dudek, Alex Dzierba, Philippe de Forcrand, Ken
Hicks, Ken-Ichi Imai, Oliver Jahn, Ambar Jain, Elizabeth Jenkins,
Kreso Kadija, Marek Karliner, Igor Klebanov, Bruce Knuteson, Vladimir
Kopeliovich, Julius Kuti, Harry Lipkin, Laurie Littenberg, Kim
Maltman, Aneesh Manohar, Wally Melnitchouk, Colin Morningstar, Takashi
Nakano John Negele, Ann Nelson, Matthias Neubert, Shmuel Nussinov,
Costas Orginos, Christoph Paus, Dan Pirjol, Michal Praszalowicz,
Claudio Rebbi, Jon Rosner, Shoichi Sazaki, Berthold Stech, Iain
Stewart, Tony Thomas, San Fu Tuan, Arkady Vainshteyn, Mark Wise,
Herbert Weigel, and Ed Witten.  I am particularly grateful to Dan
Pirjol and Herbert Weigel for comments on a draft of this paper.

This work is supported in part by the
U.S.~Department of Energy (D.O.E.) under cooperative research
agreement~\#DF-FC02-94ER40818. 
%%%%%%%%%%%%%%%%%%%%%%%%%%%%%%%%%%%%%%%%%%%%%%%%%%%%%%%%%%%%%%%%%


\begin{thebibliography}{99}
\bibitem{theory} For recent reviews of theoretical efforts to
understand the $\Theta^+$, and views, some that coincide and others
that contrast with mine, see Refs.~\cite{diak,others}.
\bibitem{jj} This section is based
on Ref.~\cite{jaffejain} where more details can be found .
\bibitem{thetawidth} Ref.~\cite{dpp} quotes a width of order 15 MeV, 
however the calculation on which that claim is based contains an 
error, 
which corrected yields an estimate of $\sim$30 MeV~\cite{widthdisc}.
\bibitem{rgr} Writing diquarks as local composite operators is a 
little more complicated than my notation suggests.  See 
Ref.~\cite{hsz}
\end{thebibliography}
\end{document}